\documentclass[prd,superscriptaddress,nofootinbib,amsmath,amssymb,aps,11pt]{revtex4-2}

\usepackage{soul}
\usepackage[justification=raggedright,singlelinecheck=false]{caption}
\usepackage{subcaption}
\usepackage{hyperref}
\usepackage{float}
\usepackage[normalem]{ulem}
\usepackage{bm}
\usepackage{amsfonts}
\usepackage{latexsym}
\usepackage[latin1]{inputenc}
\usepackage{tabularx}
\usepackage{array}
\usepackage{palatino}
\usepackage{mathpazo}
\usepackage{mathtools}
\usepackage{graphicx} 
\usepackage{appendix}
\usepackage{booktabs}
\usepackage{dcolumn}
\usepackage[dvipsnames]{xcolor}
\usepackage{multirow}
\usepackage{acro}

\DeclareAcronym{GR}{short = GR, long = General Relativity}
\DeclareAcronym{EHT}{short = EHT, long = Event Horizon Telescope}
\DeclareAcronym{ngEHT}{short = ngEHT, long = Next Generation Event Horizon Telescope Collaboration}
\DeclareAcronym{GRMHD}{short = GRMHD, long = general relativistic magneto-hydrodynamic}
\DeclareAcronym{ZAMO}{short = ZAMO, long = zero angular momentum observer}
\DeclareAcronym{EVPA}{short = EVPA, long = electric vector position angle}
\DeclareAcronym{ISCO}{short = ISCO, long = innermost stable circular orbit}
\DeclareAcronym{TCOs}{short = TCOs, long = timelike circular orbits}
\DeclareAcronym{LIGO}{short = LIGO, long = Laser Interferometer Gravitational-Wave Observatory}
\DeclareAcronym{SMBH}{short = SMBH, long = supermassive black hole}

\begin{document}
        
\title{Polarized Equatorial Emission around Kerr Black Holes with Synchronized Scalar Hair. II. Indirect images}
        
\author{Valentin O. Deliyski}
	\email{valentin.deliyski@phys.uni-sofia.bg}
	\affiliation{Department of Theoretical Physics, Faculty of Physics, Sofia University, Sofia 1164, Bulgaria}

\author{Tsanimir Angelov}
    \email{tsangelov@phys.uni-sofia.bg}
    \affiliation{Department of Theoretical Physics, Faculty of Physics, Sofia University, Sofia 1164, Bulgaria}
    
\author{Galin N. Gyulchev}
	\email{gyulchev@phys.uni-sofia.bg}
	\affiliation{Department of Theoretical Physics, Faculty of Physics, Sofia University, Sofia 1164, Bulgaria}
    \affiliation{Space Research and Technology Institute, Bulgarian Academy of Sciences, Acad. G. Bonchev Str., Bl. 1, 1113 Sofia, Bulgaria}

\author{Stoytcho S. Yazadjiev}
	\email{yazad@phys.uni-sofia.bg}	
	\affiliation{Department of Theoretical Physics, Faculty of Physics, Sofia University, Sofia 1164, Bulgaria}
	\affiliation{Institute of Mathematics and Informatics, 	Bulgarian Academy of Sciences, Acad. G. Bonchev Str., Bl. 8, Sofia 1113, Bulgaria}
    
\begin{abstract}
We investigate the intensity polarization properties of the indirect images of a geometrically and optically thin accretion disk around fully self-consistent models of Kerr black holes with synchronized scalar hair. The presence of a massive, self-gravitating scalar field can dramatically alter the lensing properties of the spacetime, resulting in morphologically complex indirect images. We show that for a representative high scalar charge model, this results in the presence of high order central ring-like images at low observer inclinations. These images are not exponentially demagnified, and their polarized intensity, relative to the direct images, is significant. We therefore conclude that the observational signature of strongly scalarized Kerr black holes may not be dominated solely by the $n=1$ image, as images with $n\geq2$ may also contribute significantly to the observed polarized emission, depending on the radial emissivity profile of the disk and the observer inclination.
\end{abstract}

\maketitle

\section{Introduction}

Radiation emitted from the vicinity of ultra-compact objects experiences strong gravitational lensing effects, which cause its trajectory to bend. This effect is strongest in spacetimes which possess a photon region\footnote{Defined as the collection of photon orbits with fixed $r$. For Kerr black holes in GR these orbits are unstable, but certain classes of solutions in modified theories of gravity can have stable orbits in the equatorial plane.}. In such cases the radiation can perform an arbitrary amount of orbits around the central object before reaching the observer. Each such half-orbit, which we label with an integer $n \ge 0$, corresponds to a separate image of the emission medium on the observer's sky. A generic property of these images in \ac{GR} is that they are exponentially de-magnified with increasing $n$, and tend towards the boundary of the compact object's shadow. At low inclinations the $n > 0$ images appear as a series of increasingly thinner concentric rings around the shadow, and are collectively referred to as the "photon ring". As the radiation completes more half-turns, it becomes a natural probe for the spacetime structure in the near vicinity of the photon region. \\

The recent discussion regarding experimentally detecting the presence of a photon ring has been prompted by the success of the \ac{EHT} Collaboration in imaging the \ac{SMBH} candidates Sgr A* and M87* \cite{EHT_M87_I, EHT_M87_VII, EHT_SgrA_I, EHT_sgrA_VIII, Moscibrodzka_2021, Tiede_2022, Paugnat_2022, Broderick_2023, Staelens_2023, PhysRevD.107.043030, Shavelle_2024, Tamar_2024} and is one of the stated goals of the \ac{ngEHT} \cite{Johnson_2023, Ayzenberg2025-qv}. These discussions focus on the photon ring, produced by black holes in \ac{GR}, for which the dominant flux contribution is expected to come from the $n = 1$ image. This is due to the exponential demagnification in $n$, which is not necessarily present when the black hole is surrounded by a self-gravitating scalar field. In such cases higher order images with $n > 1$ can have a significant contribution to the observational appearance of the photon ring - both in its intensity and polarization profile.\\

In this paper we extend our work in \cite{Deliyski2026} to look at polarized indirect images of Kerr black holes with synchronized scalar hair, which are numerical models originally developed in \cite{hairysol1, herdeiro2014, herdeiro2015, collodel2020rotating}. The above mentioned models all represent a fully self-consistent construction of black holes in massive scalar-tensor gravity, and their deviation from Kerr black holes in \ac{GR} can be characterized globally based on the normalized Noether charge q of the solution \cite{herdeiro2014}. A notable feature of these spacetimes is that due to the chaotic lensing effect for a high normalized charge $q$, the shadow region can become disconnected \cite{Cunha2015, Gyulchev2024, Wan2026}. This results in the formation of multiple disconnected images of the accretion disk for a fixed $n > 0$, which appear as offset rings on the observer's sky at low inclinations. At high inclinations the produced indirect images can have a very complicated structure as shown in \cite{Vincent_2016, Galin_2026}. In \cite{Deliyski2026} we concluded that the observable polarization pattern of the direct images is sensitive not only to global modifications to the spacetime (as measured by the normalized charge $q$), but also to local variation near the emission region, and in particular the \ac{ISCO}. In this context we study the properties of the $n > 0$ indirect images around Kerr black holes with synchronized scalar hair to determine weather the significantly more complicated image structure gives rise to reliable observational imprints of the scalar field. We follow the well established phenomenological approach of modeling synchrotron emission in the frequency range relevant to the \ac{EHT} originally outlined in \cite{Narayan_2021} for Schwarzschild black holes, later extended to Kerr black holes and arbitrary stationary axisymmetric spacetimes \cite{M_Johnson_2021, Wormhole_paper, Naked_singularity_pol, Gauss_Bonnet_pol, Tsanimir_2025, Deliyski2026} . \\

This paper is organized as follows: in section II we briefly discuss the underlying scalar-tensor theory and the classification of the numerical spacetimes we consider. In section III we discuss the phenomenological approach to modeling synchrotron emission at the \ac{EHT} observational frequencies. In section IV we present our results for two representative numerical models of scalarized black holes. Finally in section V we give a summary of our results along with a discussion.

\section{Kerr black  holes with synchronized scalar hair}

The theory that we study has two dynamical scalar fields $\varphi=(\varphi^1, \varphi^2)$, or equivalently one complex scalar field $\Psi=\varphi^{1} + i\varphi^{2}$, minimally coupled to Einstein gravity and whose action is given by 
\begin{equation}
    S=\frac{1}{4 \pi G} \int d^4 x \sqrt{-g}\left(\frac{R}{4}-\frac{1}{2} g^{\mu\nu} \delta_{a b} \partial_\mu \varphi^a \partial_\nu \varphi^b-V(\varphi)\right),
\end{equation}
with $V(\varphi)$ being the potential of the scalar fields. In the present paper we consider the standard massive potential   
\begin{equation}
    V(\varphi)=\frac{1}{2}\mu^2\psi^2
\end{equation}
where   $\mu $ is the mass of the scalar fields and $\psi^2=(\varphi^1)^2 + (\varphi^2)^2$. In order to evade the well-known scalar no-hair theorems and to preserve the stationary and axial symmetry,   the scalar fields must have the following dependence on the spacetime coordinates
\begin{equation}
    \varphi^1=\psi(r, \theta) \cos \left(\omega_s t - m \phi\right), \quad \varphi^2=\psi(r, \theta) \sin \left(\omega_s t - m \phi\right).
\end{equation}
Here $\omega_s$ has the meaning of a real-valued frequency parameter and $m$ is an azimuthal harmonic index given by an integer. The spacetime line element for these solutions is given by
\begin{equation}{\label{lineelement}}
   d s^2=-\mathcal{N} e^{2 F_0} d t^2+e^{2 F_1}\left(\frac{d r^2}{\mathcal{N}}+r^2 d \theta^2\right)+e^{2 F_2} r^2 \sin ^2 \theta\left(d \phi-\frac{\omega}{r} d t\right)^2,
\end{equation}
where $F_0,F_1,F_2$ and $\omega$ are functions which depend only on $r$ and $\theta$ while the position of the event horizon $r_H$ appears in $\mathcal{N} =1-\frac{r_H}{r}$.\\

The black hole angular frequency $\Omega_H$ and the azimuthal number must satisfy the synchronization condition $\omega_s=m\Omega_H$, which
prevents a non-zero scalar flux through the horizon. This guarantees the existence of stationary black hole solutions. The described model for Kerr black holes with synchronized scalar hair has been used in the perturbative regime to describe scalar-clouds \cite{hairysol1}, and the fully nonlinear and self-consistent numerical solutions were obtained in \cite{herdeiro2014}. The generalizations for scalar fields with a non-flat target space were presented in \cite{collodel2020rotating}. The three global charges that completely characterize  the solution are its ADM mass $M$, the total angular momentum $J$ and the Noether charge $Q$. 
The angular momentum of the scalar fields is quantized, namely $J_{\psi}=mQ$. A natural measure of   the hairiness of the black hole is the normalized the Noether charge $q=\frac{mQ}{J}=\frac{J_\psi}{J}$. The scalar clouds correspond to $q\approx0$ and models with $q\approx1$ are referred to as boson stars.

\begin{figure}[H]
    \centering
    \includegraphics[width=0.65\textwidth]{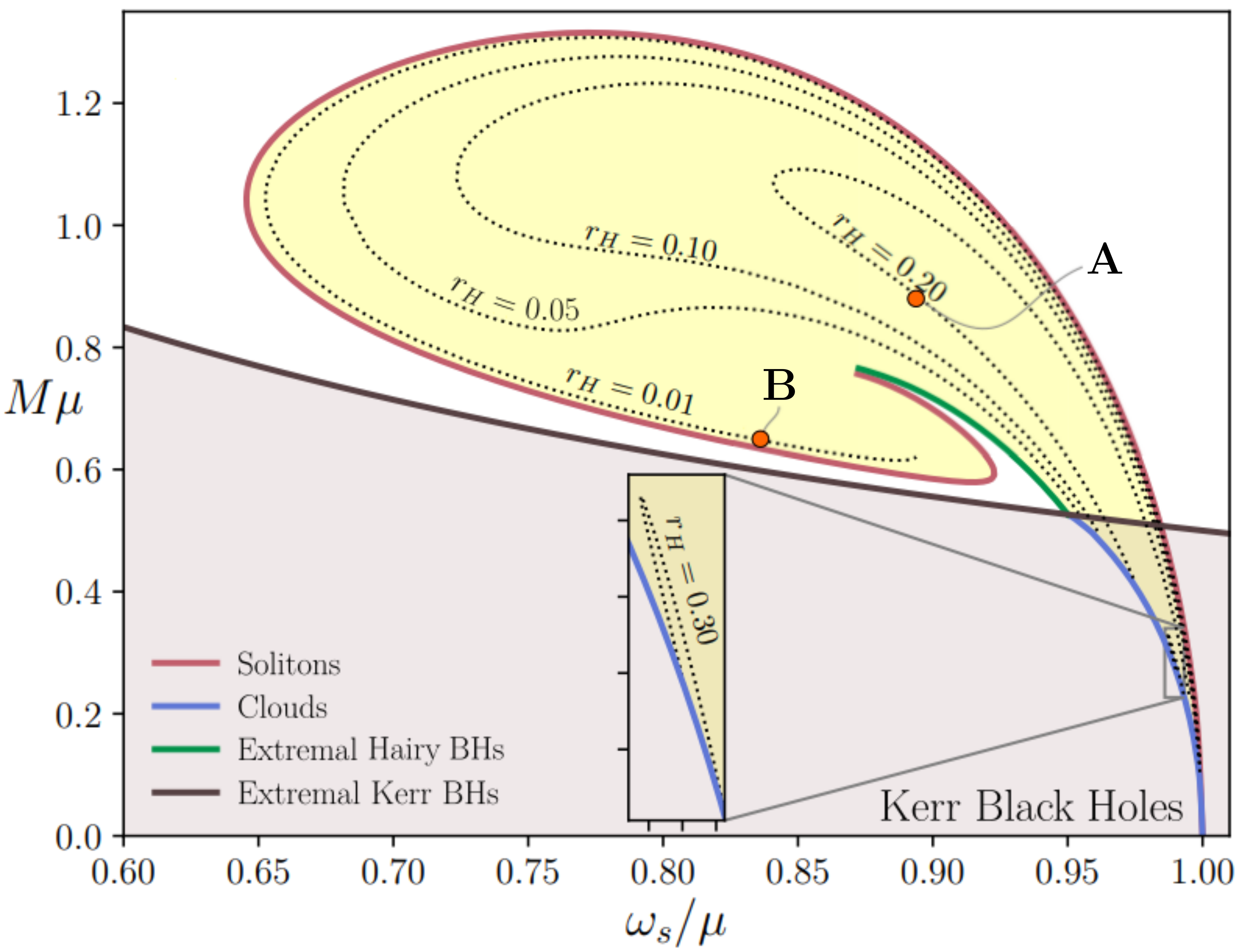}    
    \caption{\small In the $M-\omega_s$ plane, we present solution curves for fixed $r_H$ from \cite{collodel2020rotating} with $\kappa = 0$. Kerr black holes exist below the thick black line which represents extremal solutions, while hairy black holes exist in the yellow region. Orange dots indicate the solutions used for the construction of accretion disks, labeled as $\mathbf{A}$ and $\mathbf{B}$.}\label{fig:Momegaplane}
\end{figure} 

In this study we shall focus on the $m=1$ case. Considering a hairy black hole with $m=1$ the domain of existence in the $M-\omega_s$ plane is given in Fig.\ref{fig:Momegaplane}. Kerr black holes exist in the grey region below the extremal $a=M$ curve shown in black. Hairy black holes exist in the yellow region outline by the red line ($q=1, r_H=0$ - soliton region), the green line  (extremal hairy Kerr black holes with $q\neq0$) and the blue line (scalar cloud solutions with $q\approx 0$). Multiple dotted lines which represent solutions with constant horizon radii are represented within the yellow region and the black hole solutions chosen for in-depth investigation are denoted with orange dots. Each of the dotted lines begins on the Minkowski limit ($\omega_s/m=1$) and ends on the $q=0$ scalar line.

\section{Emission model}

The synchronized scalar hair, present within the studied theory, affects the propagation of polarized light, emitted from the accretion disk surrounding the black hole. To study this effect, we apply a phenomenological model first developed to analytically explain the observed polarization signatures of M87* conducted by the EHT \cite{Narayan_2021}. The model was originally developed for Schwarzschild black holes and was later generalized to the rotating Kerr black holes \cite {Johnson_2023}, 4D Gauss-Bonnet black holes \cite{Gauss_Bonnet_pol} and arbitrary spherically symmetric spacetimes \cite{Wormhole_paper}. As this study examines spacetimes with a line element in the form of \eqref{lineelement}, the following is a brief overview of the extension of the model to such general stationary and axially symmetric geometries \cite{Deliyski2026}.\\

The local orthonormal tetrad used in the analysis is that of a \ac{ZAMO}
\begin{equation}\tag{5a}
    e_{(t)}=\sqrt{\frac{g_{\phi \phi}}{g_{\phi \phi}^2-g_u g_{\phi \phi}}}\left[\partial_t-\frac{g_{\phi \phi}}{g_{\phi \phi}} \partial_\phi\right]=\frac{e^{-F_0}}{\sqrt{N}}\left[\partial_t+\frac{\omega}{r} \partial_\phi\right], 
\end{equation}
\begin{equation}\tag{5b}
    e_{(r)}=\frac{1}{\sqrt{g_{r r}}} \partial_r=\sqrt{N}e^{-F_1} \partial_r, 
\end{equation}
\begin{equation}\tag{5c}
    e_{(\theta)}=-\frac{1}{\sqrt{g_{\theta \theta}}} \partial_\theta=-\frac{e^{-F_1}}{r} \partial_\theta,  
\end{equation}
\begin{equation}\tag{5d}
    e_{(\phi)}=\frac{1}{\sqrt{g_{\phi \phi}}} \partial_\phi=\frac{e^{-F_2}}{r \sin \theta} \partial_\phi.
\end{equation}
The orthogonality conditions which relate this basis to the Minkowski metric are $g_{\mu \nu} e_{(a)}^\mu e_{(b)}^\nu=\eta_{(a)(b)}$ and $ \eta^{(a)(b)} e_{(a)}^\mu e_{(b)}^\nu=g^{\mu \nu}$. The tetrad can be used for transformations between the \ac{ZAMO} rest frame and the local rest frame of the emitting particle via the boost parameter $\vec{\beta}$\, which is related to the emission medium's four-velocity $u^{\mu}$ via
\begin{equation}
    \vec{\beta}=\frac{1}{u^{(t)}}(u^{(r)},u^{(\theta)},u^{(\phi)}), \quad u^{(a)}=\eta^{(a)(b)}e^{\mu}_{(b)}u_{\mu}.
\end{equation}
A boost along $\vec{\beta}$ of the \ac{ZAMO} tetrad is used to reach the tetrad of the rest frame of the particle $\hat{e}^{\mu}_{(a)}$
\begin{equation}
    \hat{e}^{\mu}_{(a)}={\Lambda_{(a)}}^{(b)}e^{\mu}_{(b)}.
\end{equation}
The inverse boost is used to transform components of vectors $\left({\Lambda_{(a)}}^{(b)}\right)^{-1}\equiv{\Lambda^{(b)}}_{(a)}$ (labeling $\vec{\beta}.\vec{\beta}=\beta^2$)
\begin{equation}
{\Lambda^{(b)}}_{(a)} = \begin{bmatrix}
\gamma & -\gamma\beta_r & -\gamma\beta_\theta & -\gamma\beta_\phi \\[6pt]
-\gamma\beta_r & 1 + (\gamma - 1)\frac{\beta_r^2}{\beta^2} & (\gamma - 1)\frac{\beta_r\beta_\theta}{\beta^2} & (\gamma - 1)\frac{\beta_r\beta_\phi}{\beta^2} \\[6pt]
-\gamma\beta_\theta & (\gamma - 1)\frac{\beta_\theta\beta_r}{\beta^2} & 1 + (\gamma - 1)\frac{\beta_\theta^2}{\beta^2} & (\gamma - 1)\frac{\beta_\theta\beta_\phi}{\beta^2} \\[6pt]
-\gamma\beta_\phi & (\gamma - 1)\frac{\beta_\phi\beta_r}{\beta^2} & (\gamma - 1)\frac{\beta_\phi\beta_\theta}{\beta^2} & 1 + (\gamma - 1)\frac{\beta_\phi^2}{\beta^2}
\end{bmatrix}, \quad \gamma = \frac{1}{\sqrt{1 - \beta^2}}.
\end{equation}

The emission angle $\xi$ is related to the magnetic field in the rest frame of the emitting particle $\vec{B}=\left(\hat{B}^{(r)},\hat{B}^{(\theta)},\hat{B}^{(\phi)}\right)$ and the three-momentum of the emitted photon $\vec{k}=\frac{1}{\hat{k^{(t)}}}\left(\hat{k}^{(r)},\hat{k}^{(\theta)},\hat{k}^{(\phi)}\right)$ in the following way
\begin{equation}
    \sin{\xi}=\frac{|\vec{k}\times\vec{B}|}{|\vec{B}|}.
\end{equation}

Synchrotron emission depends strongly on the particle pitch angle, relative to the magnetic field. In the ultra-relativistic regime $\gamma >> 1$, synchrotron emission is confined to a cone along the local 3-velocity of the particle, with an opening angle $\approx \gamma^{-1}$. This allows us to identify the pitch angle, governing the emission process, with the angle $\xi$. The specific intensity has the following dependence on $\xi$ in the different regimes depending on the temperature of the particles $T$
\begin{equation}
    I_{\nu}\propto\sin^{2/3}\xi,\quad h\nu\ll kT,
\end{equation}
\begin{equation}
    I_{\nu}\propto e^{x^{-1/3}},x\propto \frac{1}{\sin{\xi}}\quad h\nu\gg kT.
\end{equation}
Further details and formulas on how to bridge the two regimes can be found in the literature \cite{Mahadevan_1996, Marszewski_2021}. We chose a dependence between the specific intensity and the emission angle which is in accordance with the results obtained from observations of M87* \cite{Narayan_2021}. The general relativistic magneto-hydrodynamic models for this black hole give the dependence $I_{\nu}\propto\nu^{-\alpha_{\nu}}\rightarrow I_{\nu}\propto\sin^{1+\alpha_{\nu}}\xi$. At the 230 GHz, which is the current observational frequency of \ac{EHT},
$I_{\nu}\propto \sin^2\eta$ which validates the choice of $\alpha_{\nu} = 1$ that we adopt in our work.\\

The emitted radiation's polarization vector has a spatial part $\vec{f}$ which is given by the following expression in the rest frame of the emitting particle
\begin{equation}
    \vec{f}=\frac{\vec{k}\times\vec{B}}{|\vec{B}|}.
\end{equation}
Furthermore we use the gauge freedom to choose $\hat{f}^{(t)}=0$. The polarization vector then satisfies
\begin{equation}
    \hat{f}^{(a)}\hat{f}_{(a)}=\sin^2\xi.
\end{equation}
The task of constructing polarized images of the accretion disk then reduces to parallel transporting $f^{\mu}$ from the emission point to the observer along null geodesics, subject to the constraint $k^\mu f_\mu=0$. We do this numerically by solving the following system of ODEs
\begin{equation}
    k^{\nu}\nabla_{\nu}k^{\mu}=0,
\end{equation}
\begin{equation}
    k^{\nu}\nabla_{\nu}f^{\mu}=0,
\end{equation}
whose solutions at the observer we use to define the two main observables - the specific intensity and the \ac{EVPA}:
\begin{equation}
    I_{\nu}\propto \delta^{3+\alpha_{\nu}} \ell_p \sin^{1+\alpha_\nu}\xi, 
\end{equation}
\begin{equation}
    EVPA=\arctan\left(-\frac{f_{obs}^{(\phi)}}{f_{obs}^{(\theta)}}\right),
\end{equation}
where $\delta$ denotes the Doppler factor $\frac{u^{\mu}k_{\mu}|_{obs}}{u^{\mu}k_{\mu}|_{emitter}}$ and $\ell_p = \frac{\hat{k}^{(t)}}{\hat{k}^{(\theta)}}H$ is the projected disk thickness $H$ along the ray path. Given that the model specifies $I_\nu$ up to a multiplicative factor (see the discussion in \cite{Deliyski2026}), we can set $H = 1$. The emitting particle is further assumed to be co-rotating with the black hole on a timelike circular geodesic in its equatorial plane. This means that the emitter's four-velocity is given by
\begin{equation}
    u^{\mu}_\text{emitter}=u^t(1,0,0,\Omega),
\end{equation}
where $\Omega$ is the particle's angular velocity and $u^t$ is determined from the normalization condition $u_{\mu}u^{\mu}= - 1$.
\begin{equation}
    \Omega = \frac{-\partial_r g_{t\phi} + \sqrt{\partial_r g_{t\phi}^2 - \partial_r g_{tt} \partial_r g_{\phi\phi}}}{\partial_r g_{\phi\phi}}, \quad u^t = \frac{1}{\sqrt{-g_{tt} - 2g_{t\phi}\Omega - g_{\phi\phi}\Omega^2}}.
\end{equation}

\section{Polarized indirect images at low inclinations}

Realistic equilibrium models of magnetized accretion disks show that the region of maximum density (where we expect to have significant emission) tends to lie near the \ac{ISCO}, even though the inner edge of the disk can extend below it \cite{Gimeno_2019, Gimeno_2021}. We thus concentrate our analysis on the \ac{ISCO} and its immediate vicinity. As in \cite{Deliyski2026}, we present our results by selecting two representative models: \textbf{A}, with a low normalized charge, and \textbf{B}, with a high normalized charge, corresponding to models $\mathbf{V}_{\,0.2}$ and $\mathbf{II}_{\,0.01}$, respectively, in \cite{Deliyski2026}. Their physical properties are summarized in Table \ref{tab_sol_properties} of the Appendix. \\

 In order to compare the intensity variation across all images of a given orbit in a unified way throughout this paper, we parametrize the apparent shape of the images with their natural parameter $s$. We call this the "normalized length" across the image. The yellow triangles on all figures correspond to the point $s = 0$, the yellow dots to the maximum emission, and the arrows indicate the direction of increasing $s$. We label the indirect images with roman numerals (in cases where more than one image of the accretion disk is present), starting from the outermost one. The third panel on all figures shows the intensity of the indirect images, normalized to the maximum of the corresponding direct image.

\subsection{Low normalized charge model at low observer inclinations}

\begin{figure}[H]
    \centering
    \includegraphics[width=0.8\textwidth]{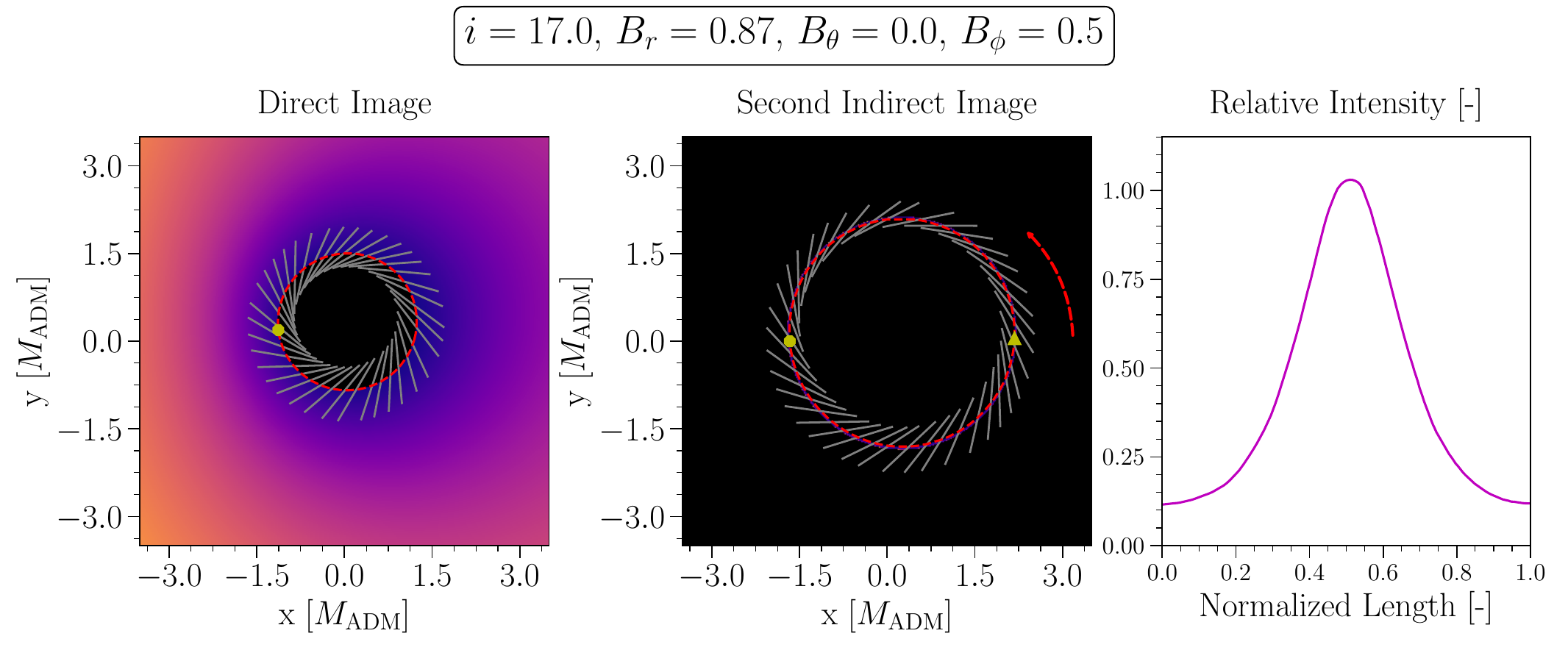}     
    \caption{\small Comparison between the direct and $n = 2$ indirect image of the \ac{ISCO} of model \textbf{A} for an equatorial magnetic field and a $17^\circ$ observer inclination. The rightmost panel shows the intensity across the images.}\label{fig:V_isco_17_deg_n2}
\end{figure}

We first analyze the representative low normalized charge model \textbf{A} at low inclinations. Its lensing properties are qualitatively similar to that of Kerr black holes. Namely for each $n$ there is exactly one image of the accretion disk, and the $n > 1$ images are exponentially demagnified with respect to the $n = 0$ and $n = 1$ ones (see figure \ref{fig:V_isco_17_deg_n2}). This demagnification is a consequence of the Kerr-like structure of the model's null geodesic structure. It also results in the polarization pattern of the higher order images retaining certain properties, characteristic of the photon ring around Kerr black holes in GR \cite{Universal}. In particular the property $\vec{f}_{n+2} = \vec{f}_n$, which can approximately be seen in figure \ref{fig:V_isco_17_deg_n2} (even though the equality holds strictly in the limit $n \rightarrow \infty$).\\

\begin{figure}[H]
    \centering
    \includegraphics[width=0.8\textwidth]{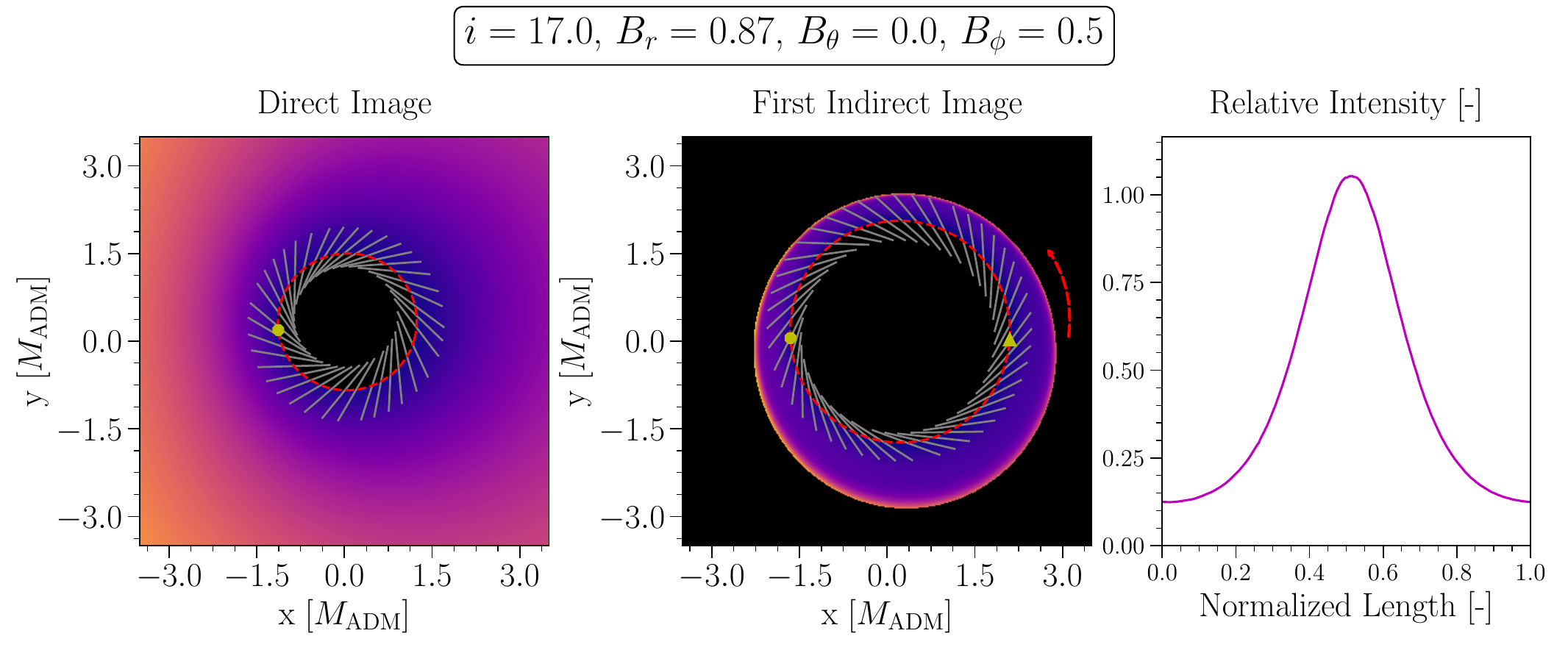}    
    \includegraphics[width=0.8\textwidth]{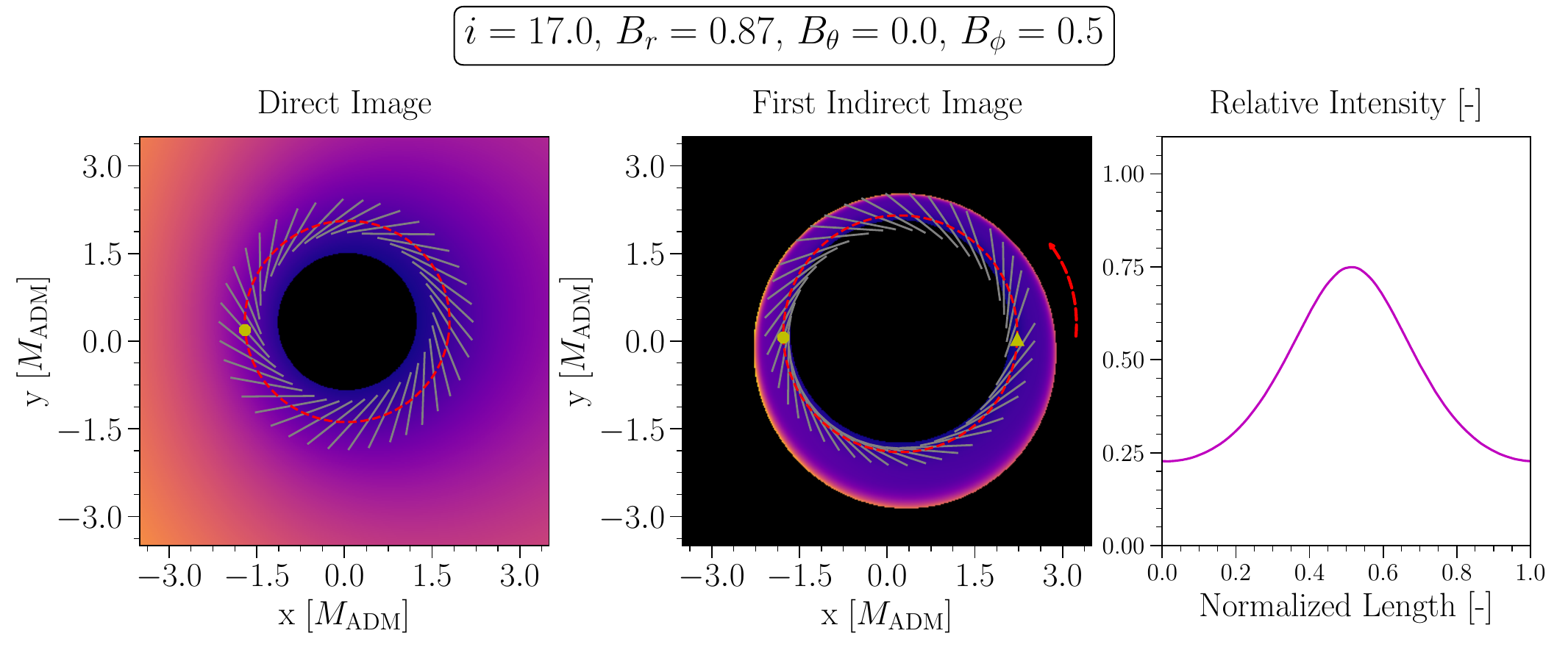}    
    \caption{\small Comparison between the direct and $n = 1$ indirect images of the $r = r_\text{ISCO}$ and$r = 2r_\text{ISCO}$ orbits of model \textbf{A} for an equatorial magnetic field and a $17^\circ$ observer inclination. The rightmost panel shows the intensity across the images.}\label{fig:V_1_2_isco_17_deg_n1}
\end{figure}

Furthermore, the thin accretion disk which we consider is axially symmetric. When we consider an equatorial magnetic field, the emission medium can also be taken to be reflection symmetric across the equator. In such cases \cite{Universal} show that in the large $n$ limit the covariant polarization vector satisfies $f_{n+1}^\theta\approx - f^\theta_n$ for an equatorial source. The observational consequence of this symmetry is that the polarization pattern approximately flips across $y = 0$ on the observers screen between successive images in $n$. We can observe this effect between the $n = 0, 1 \text{ and } 2$ images on figures \ref{fig:V_1_2_isco_17_deg_n1} and \ref{fig:V_isco_17_deg_n2}.
The exponential demagnification also means that the observed flux distribution and polarization pattern of the photon ring will be dominated by those of the $n = 1$ image.

\subsection{Low normalized charge model at high observer inclinations}

\begin{figure}[H]
    \centering
    \includegraphics[width=0.85\textwidth]{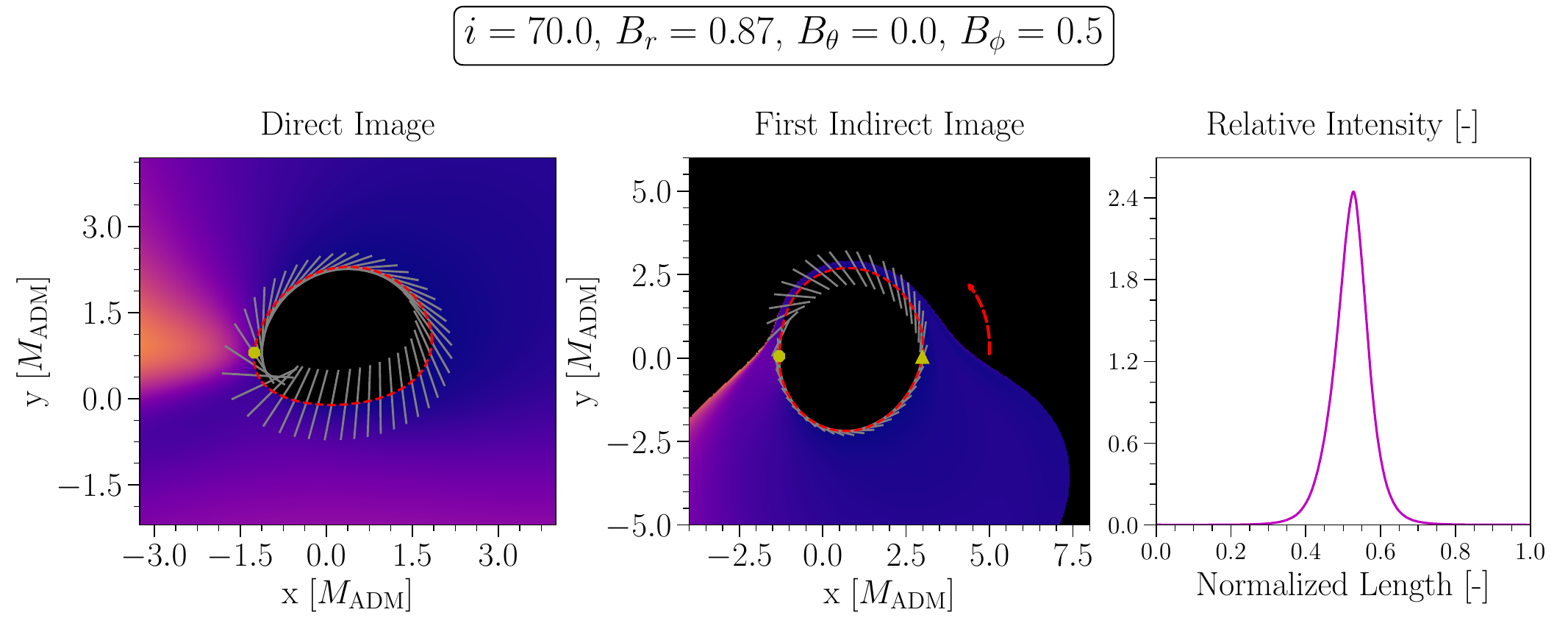} 
    \includegraphics[width=0.85\textwidth]{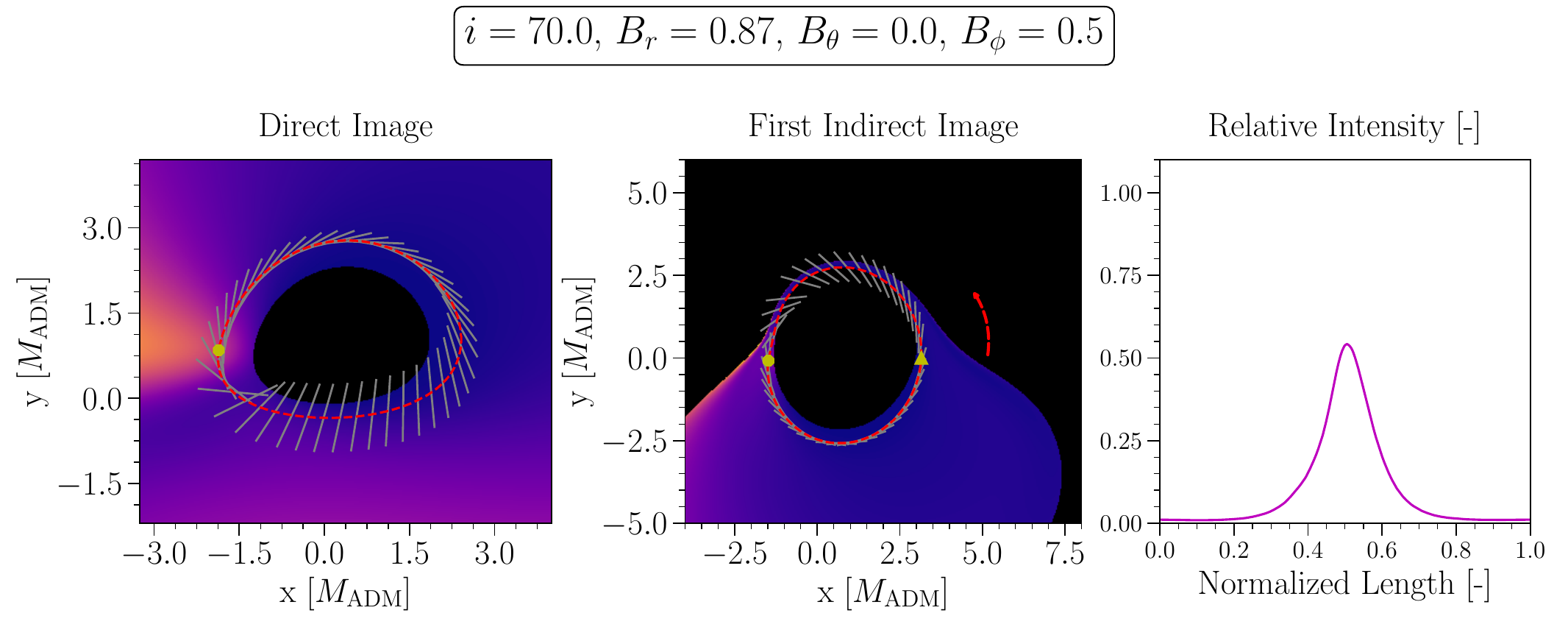} 
    \caption{\small Comparison between the direct and $n = 1$ indirect images of the $r = r_\text{ISCO}$ and $r = 2r_\text{ISCO}$ orbits of model \textbf{A} for an equatorial magnetic field and a $70^\circ$ observer inclination. The rightmost panel shows the intensity across the images.}\label{fig:V_1_2_isco_70_deg_n1}
\end{figure}

The high inclination case does not show any qualitatively new effects. The intensity of the indirect images is high compared to the direct one in the vicinity of the \ac{ISCO}, and falls off quickly with increasing orbital radius (visible on figure \ref{fig:V_1_2_isco_70_deg_n1}).

The result is that the indirect images again form the characteristic intense ring-like image on the sky, dominated by the flux distribution and polarization pattern of the $n = 1$ image.

\subsection{High normalized charge model at low observer inclinations}

\begin{figure}[H]
    \centering
    \includegraphics[width=1\textwidth]{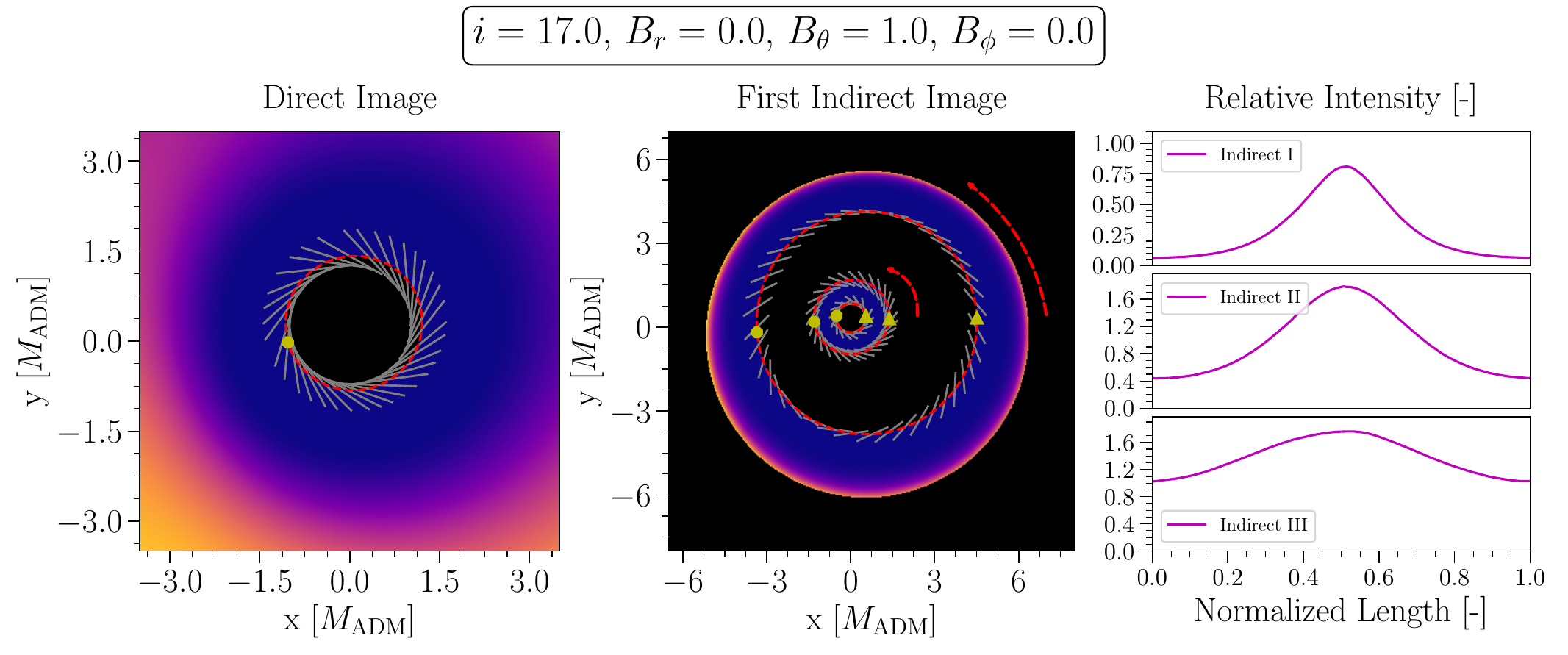}    
    \includegraphics[width=1\textwidth]{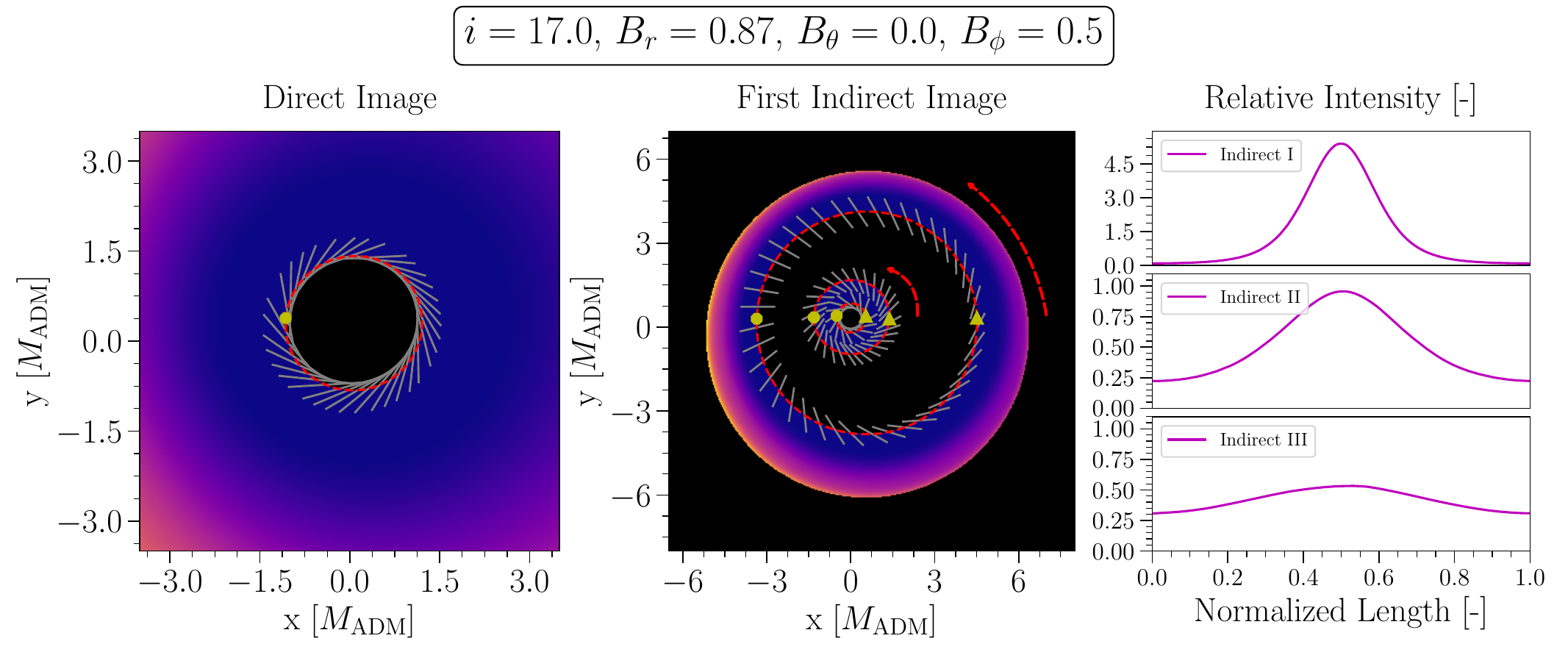}    
    \caption{\small Comparison between the direct and $n = 1$ indirect image of the \ac{ISCO} orbit of model \textbf{B} for a vertical and equatorial magnetic field and a $17^\circ$ observer inclination. The rightmost panel shows the intensity across the images.}\label{fig:II_isco_17_deg}
\end{figure}

We now present our results for the representative high normalized charge model \textbf{B}. In figure \ref{fig:II_isco_17_deg} we present the $n = 1$ image of a thin accretion disk, viewed at a $17^\circ$ inclination. Unlike the model \textbf{A}, this high normalized charge model produces indirect images that differ significantly from an analogous Kerr black hole in GR. In particular we see the formation of a central ring structure, which contains two separate \ac{ISCO} images (labeled images II and III). In order to be observable under realistic experimental conditions, the relative intensity of these images compared to the direct one must be on the order of magnitude of unity. We can see from the rightmost panel of \ref{fig:II_isco_17_deg} that while this ratio depends strongly on the magnetic field geometry, it is nevertheless close to unity across a significant portion of the image.

\begin{figure}[H]
    \centering
    \includegraphics[width=1\textwidth]{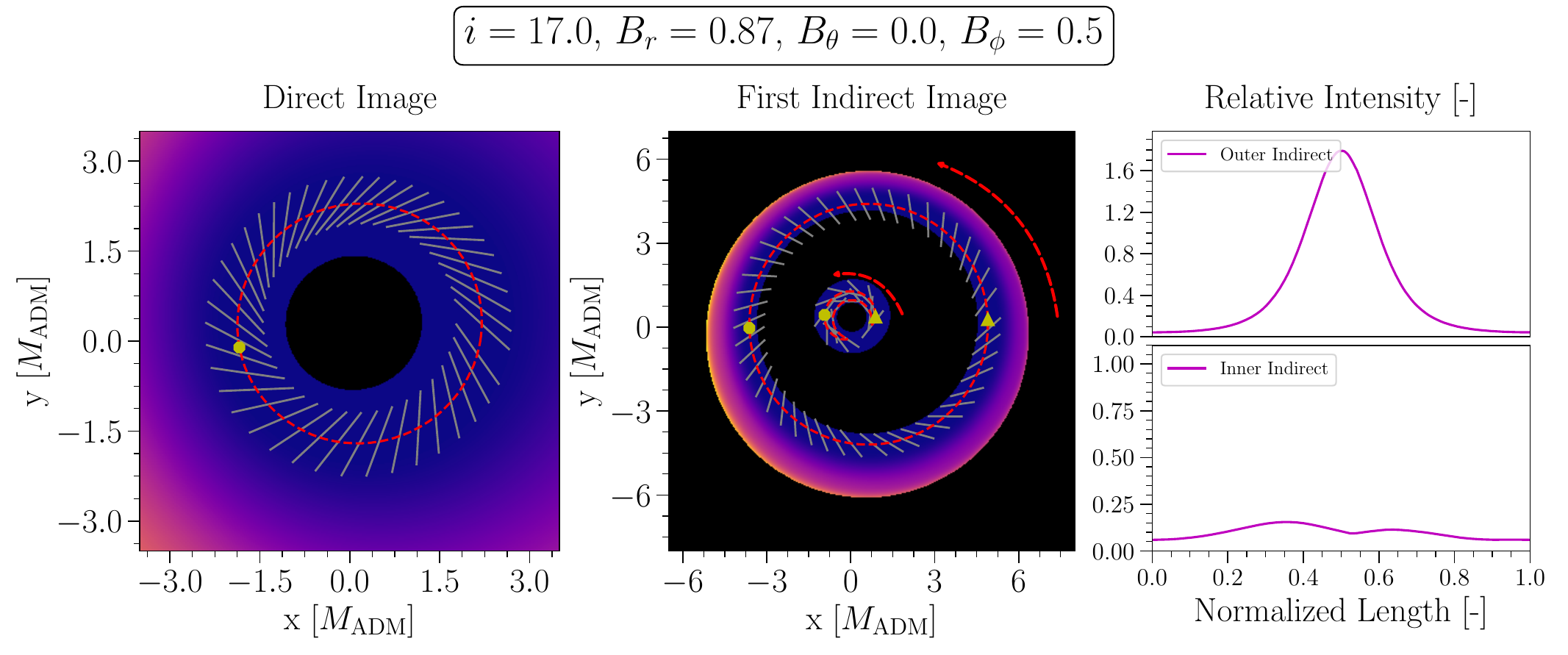}  
    \caption{\small Comparison between the direct and $n = 1$ indirect image of the $r = 1.8r_\text{ISCO}$ orbit of model \textbf{B} for a vertical magnetic field and a $17^\circ$ observer inclination. The rightmost panel shows the intensity across the images. The $r = 1.8r_\text{ISCO}$ value is the approximate orbital radius at which the two inner images from figure \ref{fig:II_isco_17_deg} combine.}\label{fig:II_1.8_isco_17_deg}
\end{figure}

This image structure changes drastically for higher emission radii. In figure \ref{fig:II_1.8_isco_17_deg} we show the polarized image of the $r = 1.8r_\text{ISCO}$ image for model \textbf{B}. We find that for higher orbital radii, the two innermost images combine into a single arc-like one\footnote{Similar images have been shown to result from geodesics which encounter a potential barrier in their effective potential and reflect back towards the observer \cite{Gyulchev2019, Gyulchev2020, Gyulchev2021} or ones that pass through wormhole throats \cite{Paul_2020, Rahaman2021}.} , whose intensity is significantly lower than that of the direct image. This behavior is similar across all magnetic field configurations and implies that the observational signature of these central images will depend strongly on the radial emission profile in the near vicinity of the \ac{ISCO}.

\begin{figure}[H]
    \centering
    \includegraphics[width=1\textwidth]{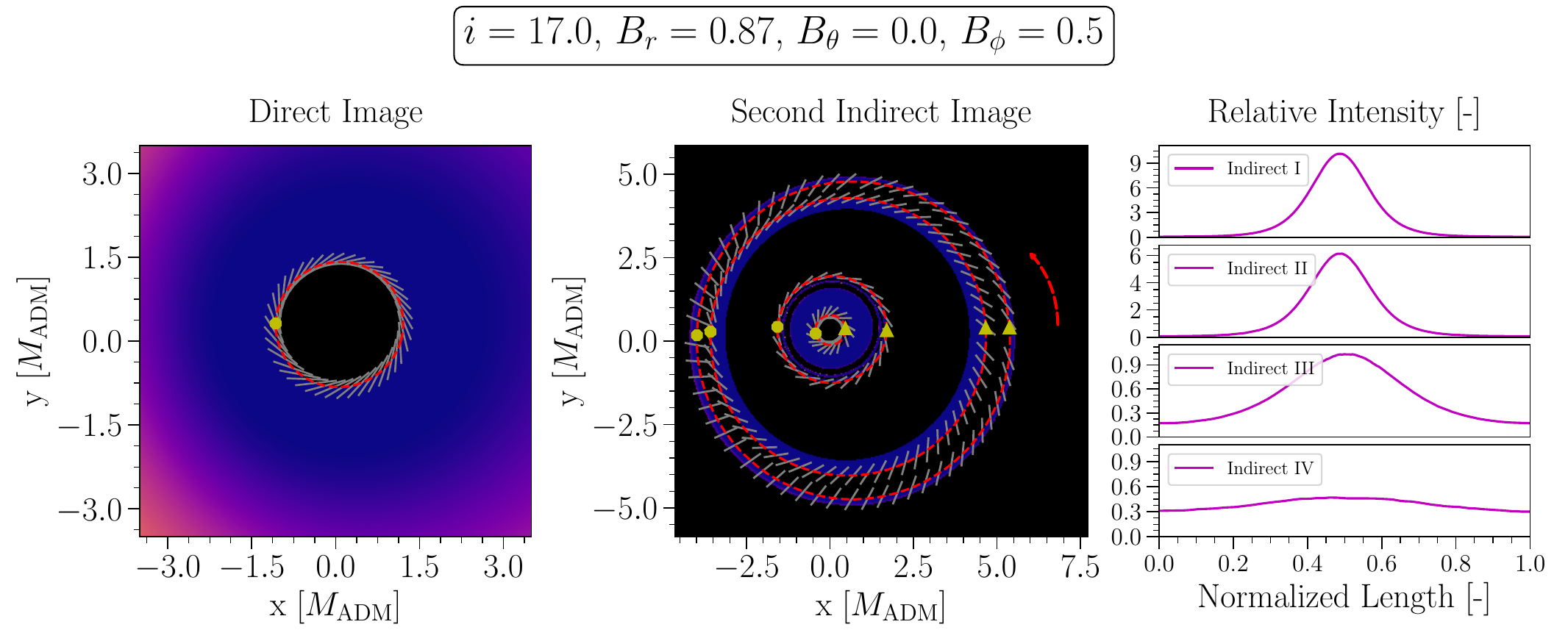}  
    \caption{\small Comparison between the direct and $n = 2$ indirect image of the $r = r_\text{ISCO}$ orbit of model \textbf{B} for a vertical magnetic field and a $17^\circ$ observer inclination. The rightmost panel shows the intensity across the images.}\label{fig:II_1_isco_17_deg_order_2}
\end{figure} 

Further notable features of these indirect images can be found in the $n = 2$ case, shown on figure \ref{fig:II_1_isco_17_deg_order_2}. Firstly, we find that the vicinity of the \ac{ISCO} region is mapped to four disconnected ringlike images on the observer's sky. Secondly, most of these images have a significant intensity, relative to the direct image. We find that this result also holds for higher orbits (although indirect image III's intensity becomes negligible at around $r \approx 2r_\text{ISCO}$. Lastly, and most importantly from an observational point of view, these images are not exponentially de-magnified as in the GR Kerr black hole case. This means that the dominant contribution to the photon-ring flux in model \textbf{B} does not necessarily originate solely from the $n=1$ image. 

The multiplicity of the indirect images reflects the nontrivial structure of null geodesics in the strongly scalarized spacetime. Unlike the Kerr geometry, where null geodesic motion is completely integrable, the more general stationary and axisymmetric spacetime considered here may exhibit a considerably more intricate geodesic structure, giving rise to multiple spatially disconnected images of the same equatorial emitting orbit.

This can have serious ramifications for the observed polarization pattern. For example, indirect images I and II on figure \ref{fig:II_1_isco_17_deg_order_2} are very close to each other on the sky, while their projected polarization vectors $\vec{f}$ are nearly orthogonal in the region of higher intensity. Observationally this could lead to a polarization "scrambling" effect at finite resolutions.\\

The potential to detect such central images with current and next generation \ac{EHT} observations has been recently studied in \cite{Olivares2020, Eichhorn_2023_I, Eichhorn_2023_II, deKleuver2024, PhysRevD.111.064068}. They show that with the capabilities of the 2017 \ac{EHT} array, thin central images are not resolved, but still leave an imprint in the central brightness depression. This imprint is characterized by a depression ratio $\hat{f}_c$, defined as \cite{Akiyama_2019}:
\begin{equation}
    \hat{f}_c = \frac{\text{Minimum flux in the depression}}{\text{Mean flux in the "ring"}}.
\end{equation}
Here the "ring"\footnote{The term "ring" in the context of reconstructed images from the \ac{EHT} should not be confused with the photon ring as defined in the introduction. The former refers to the apparent shape of the entire accretion disk, as realistically seen by the \ac{EHT}, while the latter refers to an idealized infinite resolution observation of all $n \ge 1$ images of the disk.} refers the observational appearance of the entire disk (with images for all $n$ present). With the current capabilities of the \ac{EHT}, this is expected to have a ring-like shape with a central brightness depression. In the context of the images of figures \ref{fig:II_isco_17_deg}, \ref{fig:II_1.8_isco_17_deg} and \ref{fig:II_1_isco_17_deg_order_2}, the brightness depression will have its mean value increased by the presence of the central images, and the polarization properties of the "ring" will have a significant contribution from the outer $n = 1$ and $n = 2$ images.\\

Future observations with the planned \ac{ngEHT} will still not have the necessary angular resolution to preserve the morphological appearance of these central images, but their imprint will become significantly more pronounced. In particular, they can lead to the appearance of localized peaks in the central depression, which were shown to reach a maximum flux of $\approx 15 - 30 \%$ of the main "ring" for certain models of naked singularities \cite{PhysRevD.111.064068}.

\subsection{High normalized charge model at high observer inclinations}

\begin{figure}[H]
    \centering
    \includegraphics[width=1\textwidth]{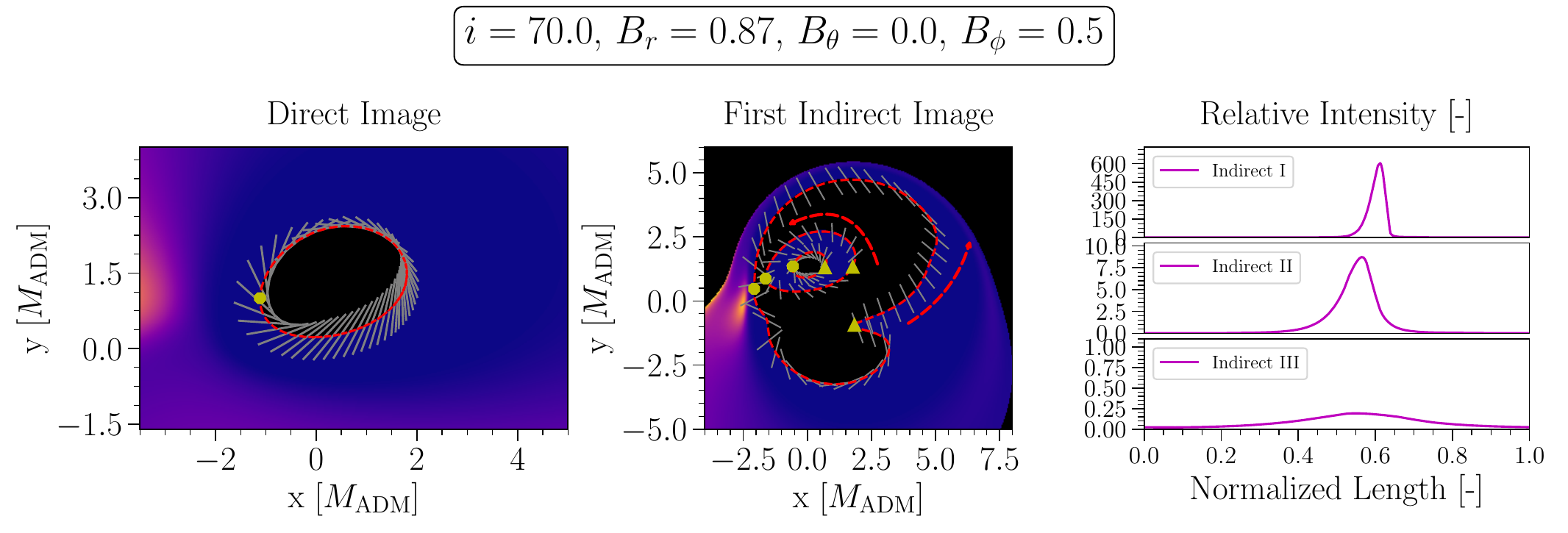}    
    \includegraphics[width=1\textwidth]{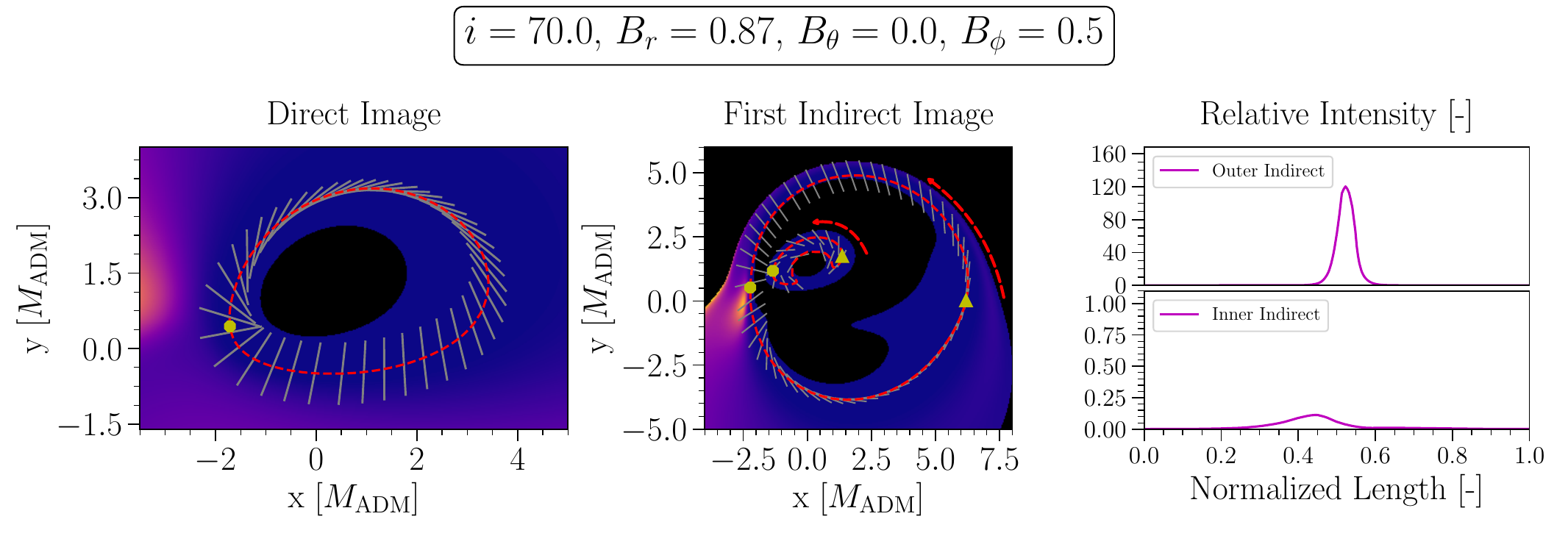}    
    \caption{\small Comparison between the direct and $n = 1$ indirect images of the $r = r_\text{ISCO}$ and $r = 1.6r_\text{ISCO}$ orbits of model \textbf{B} for an equatorial magnetic field and a $70^\circ$ observer inclination. The rightmost panel shows the intensity across the images. The $r = 1.6r_\text{ISCO}$ value is the approximate orbital radius at which the two inner images from the top middle panel combine.}\label{fig:II_1_1.6_isco_70_deg}
\end{figure}

The high inclination case retains the central images discussed in the previous section, while drastically deforming their apparent shape. On figure \ref{fig:II_1_1.6_isco_70_deg} we present the $n = 1$ indirect image of the $r = r_\text{ISCO}$ and$r = 1.6r_\text{ISCO}$ orbits. A notable feature of the \ac{ISCO} image is the concave region around $s = 0$. From the bottom panels of figure \ref{fig:II_1_1.6_isco_70_deg} we can see that this region only maps the near vicinity of the \ac{ISCO} and (from the intensity panels) that its emission is strongly suppressed by the combined gravitational and kinematic redshift at the receding side of the disk. It is then reasonable to assume that while this region of the image is a striking feature of the model, it is likely an observationally unimportant one.

\begin{figure}[H]
    \centering
    \includegraphics[width=1\textwidth]{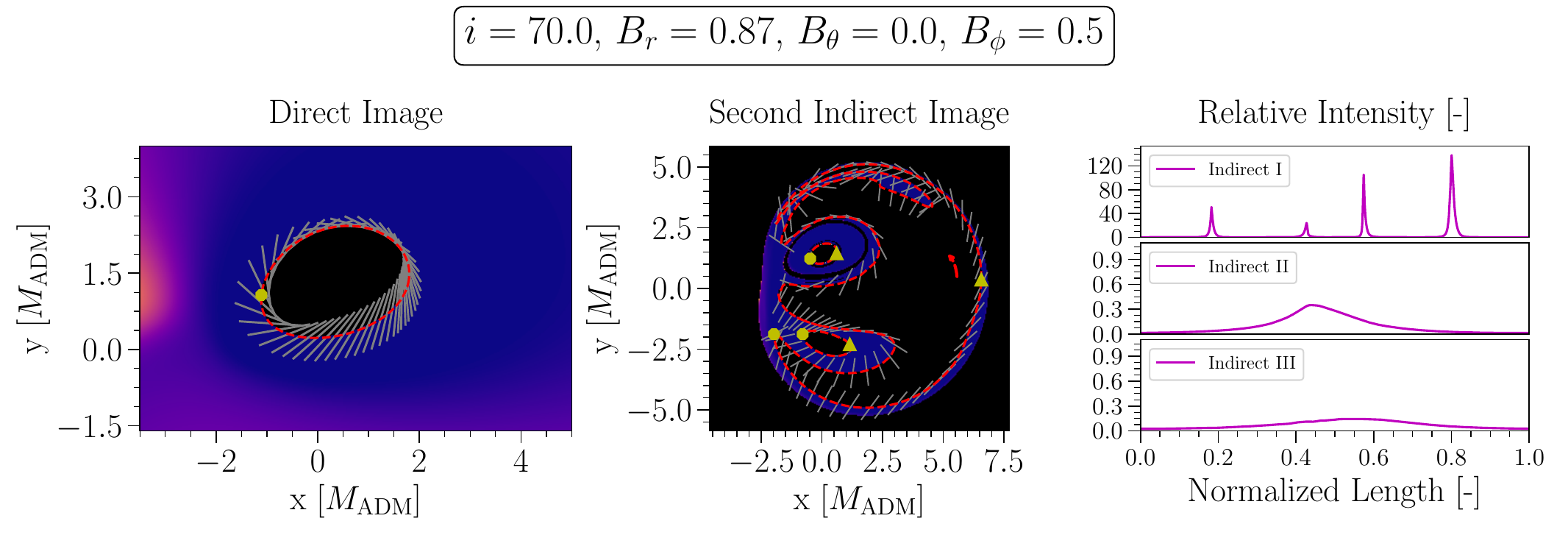}    
    \caption{\small Comparison between the direct and $n = 2$ indirect images of the $r = r_\text{ISCO}$ orbit of model \textbf{B} for an equatorial magnetic field and a $70^\circ$ observer inclination. The rightmost panel shows the intensity across the images.}\label{fig:II_1_isco_70_deg_n2}
\end{figure} 

The dependence of the intensity profile of the $n = 1$ images on the emission radius is similar to the low-inclination case. The bottom panel on figure \ref{fig:II_1_1.6_isco_70_deg} shows that at an emission radius $ r = 1.6 r_\text{ISCO}$ the inner indirect image has a negligible intensity, compared to the direct one. We also observe that the two images from the top panel again merge into a single concave one, as in the low inclination case.\\

The polarization pattern of the outermost $n = 1$ \ac{ISCO} image also exhibits a more complicated twisting, while the two inner ones retain a twisting similar to their the low inclination counterparts. As the emission radius increases, the outermost $n = 1$ image's polarization pattern starts to follow the expected Kerr-like relationship to the direct image.\\

A significantly more complicated image structure presents itself for the $n = 2$ indirect images, presented on figure \ref{fig:II_1_isco_70_deg_n2}. We identify three separate \ac{ISCO} images, the outermost of which (labeled as image "I") has a nontrivial winding pattern around the origin and presents four peaks in its intensity profile, two of which are $\approx 20\%$ of the outer $n = 1$ image's relative peak. The corresponding position of these peaks on the image plane is again labeled with a yellow dot. Indirect image $II$ has a peak intensity of $\approx 30\%$ the value of the direct image, while the final (image $III$) one has a negligible relative intensity.\\

It is also interesting to note that the peak relative intensity of the $n = 2$ image is $\approx 120$, while that of the $n = 1$ is $\approx 600$. This is in contrast with the low inclination case, where the peak of the $n = 2$ image is larger than that of the $n = 1$ by a factor of $\approx 2$. The observational implications of this (for the particular model \textbf{B}) are that at high inclinations, the dominant flux contribution of the higher order images will come from the $n = 1$ one. This renders the $n \ge 2$ images a subdominant effect, even though they have a large apparent size on the observer's screen.

\section{Conclusion}

Our results show that the observational signatures of indirect images for weakly scalarized Kerr black holes remain phenomenologically similar to Kerr black holes in \ac{GR}. This result is unsurprising, as these models can be viewed as a perturbation of a black hole with no scalar charge. The photon region remains qualitatively similar, which results in polarized images that exhibit generic properties characteristic of black holes in \ac{GR}. Namely the relation between the polarization vector for successive indirect images, outlined in \cite{Universal}, and their exponential demagnification with increasing $n$. We would thus expect the dominant observation signal of these models' photon ring to come from the $n = 1$ image, just as for black holes in \ac{GR}.\\

This changes drastically when one considers models with a high scalar charge. These models are globally closer to boson stars than pure black hole solutions, and their lensing properties differ substantially. At low inclinations, we observe the formation of central ring-like and cusp-like images of the emitting region near the \ac{ISCO}. Due to the more complicated geodesic structure of these models, the resulting indirect images are not exponentially demagnified in $n$. If a significant portion of these images emit with an intensity, comparable to that of the $n = 0$ and $n = 1$ images, they will be observationally relevant. We show that at low inclinations, the intensity ratio of the $n = 2$ images and the $n = \{0, 1\}$ ones can reach up to $\approx 900\%$ for the \ac{ISCO} orbit, and remains significant out until $r\approx 2r_\text{ISCO}$. Depending on the radial flux distribution of the disk, it is possible that the $n = 2$ images contribute significantly to the observational signal of the photon ring.\\

At high inclinations our model with a high scalar charge produces very morphologically complicated $n = 2$ images. Unlike the low inclination case, their relative intensity to the $n = 1$ images is at most $\approx 20\%$ for the \ac{ISCO} image, and further decrease for higher orbital radii. The regions of non-negligible intensity are also concentrated in very narrow regions on the observer's sky, and would likely be drowned out by the $n = 1$ signal. We can therefore conclude that for the representative high scalar charge model we consider, high inclination $n \ge 2$ images are not observationally significant. \\

We should note that while our results hold strictly for the particular models we have chosen, this is enough to establish the principle that spacetimes with more complicated lensing properties allow for observationally significant images with $n\ge 2$.

\begin{appendix}

\section{Physical quantities of the considered solutions}\label{Appendix Solutons}

Below in table \ref{tab_sol_properties} we provide a detailed description of the physical properties of the solutions, considered in this paper. They are taken from \cite{collodel2020rotating} in the case of a flat target space metric. The labels in the first column coincide with the labels in figure \ref{fig:Momegaplane}.

\begin{table}[H]
  \centering
  \footnotesize
    \setlength{\tabcolsep}{7.7pt} % Default value: 2pt
    \renewcommand{\arraystretch}{1.4} % Default value: 1  
    \vspace{1.5mm} 
  \begin{tabular}{cccccccccccc}
    \hline\hline
    Model & $\omega_{s}/\mu$ & $M\mu$ & $J\mu^2$ & $M_{BH}\mu$ & $J_{BH}\mu^2$ & $M_{\psi}\mu$ & $J_{\psi}\mu^2$ & $J_{\psi}/J$ & $J/M^2$ & $J_{BH}/M_{BH}^2$ \\
    \hline
    $\mathbf{A}$   & 0.8955 & 0.8787 & 0.6790 & 0.2796 & 0.1012 & 0.5990 & 0.5776 & 0.8508 & 0.8793 & 1.2946 \\
    $\mathbf{B}$   & 0.8353 & 0.6482 & 0.4068 & 0.0049 & 0.0011 & 0.6434 & 0.4057 & 0.9973 & 0.9680 & 46.444 \\    
    \hline\hline
  \end{tabular}
  \caption{\label{tab_sol_properties}\footnotesize Physical properties of the hairy black hole solutions, considered in this study. Model \textbf{A} corresponds to model $\mathbf{V}_{\,0.2}$ from \cite{Deliyski2026}, while model \textbf{B} corresponds to $\mathbf{II}_{\,0.01}$.}   
\end{table}

\end{appendix}

\section*{Acknowledgments}
This study is financed by the Bulgarian National Science Fund (NSF) under Grant KP-06-DV/8 within the funding programme ``VIHREN--2024''. 

\bibliography{References2}

@ARTICLE{Ayzenberg2025-qv,
  title     = "Fundamental physics opportunities with the next-generation Event
               Horizon Telescope",
  author    = "Ayzenberg, Dimitry and Blackburn, Lindy and Brito, Richard and
               Britzen, Silke and Broderick, Avery E and Carballo-Rubio,
               Ra{\'u}l and Cardoso, Vitor and Chael, Andrew and Chatterjee,
               Koushik and Chen, Yifan and Cunha, Pedro V P and Davoudiasl,
               Hooman and Denton, Peter B and Doeleman, Sheperd S and Eichhorn,
               Astrid and Eubanks, Marshall and Fang, Yun and Foschi, Arianna
               and Fromm, Christian M and Galison, Peter and Ghosh, Sushant G
               and Gold, Roman and Gurvits, Leonid I and Hadar, Shahar and
               Held, Aaron and Houston, Janice and Hu, Yichao and Johnson,
               Michael D and Kocherlakota, Prashant and Natarajan, Priyamvada
               and Olivares, H{\'e}ctor and Palumbo, Daniel and Pesce, Dominic
               W and Rajendran, Surjeet and Roy, Rittick and {Saurabh} and
               Shao, Lijing and Tahura, Shammi and Tamar, Aditya and Tiede,
               Paul and Vincent, Fr{\'e}d{\'e}ric H and Visinelli, Luca and
               Wang, Zhiren and Wielgus, Maciek and Xue, Xiao and Yakut, Kadri
               and Yang, Huan and Younsi, Ziri",
  journal   = "Living Rev. Relativ.",
  publisher = "Springer Science and Business Media LLC",
  volume    =  28,
  number    =  1,
  month     =  jun,
  year      =  2025,
  copyright = "https://creativecommons.org/licenses/by/4.0",
}

@article{Eichhorn_2023_I,
doi = {10.3847/1538-4357/accced},
url = {https://doi.org/10.3847/1538-4357/accced},
year = {2023},
month = {jun},
publisher = {The American Astronomical Society},
volume = {950},
number = {2},
pages = {117},
author = {Eichhorn, Astrid and Gold, Roman and Held, Aaron},
title = {Horizonless Spacetimes As Seen by Present and Next-generation Event Horizon Telescope Arrays},
journal = {The Astrophysical Journal}
}

@article{PhysRevD.111.064068,
  title = {Observing naked singularities with the present and next-generation Event Horizon Telescope},
  author = {Deliyski, Valentin and Gyulchev, Galin and Nedkova, Petya and Yazadjiev, Stoytcho},
  journal = {Phys. Rev. D},
  volume = {111},
  issue = {6},
  pages = {064068},
  numpages = {25},
  year = {2025},
  month = {Mar},
  publisher = {American Physical Society},
  doi = {10.1103/PhysRevD.111.064068},
  url = {https://link.aps.org/doi/10.1103/PhysRevD.111.064068}
}

@article{Eichhorn_2023_II,
doi = {10.1088/1475-7516/2023/01/032},
url = {https://doi.org/10.1088/1475-7516/2023/01/032},
year = {2023},
month = {jan},
publisher = {IOP Publishing},
volume = {2023},
number = {01},
pages = {032},
author = {Eichhorn, Astrid and Held, Aaron},
title = {Quantum gravity lights up spinning black holes},
journal = {Journal of Cosmology and Astroparticle Physics}
}

@article{deKleuver2024,
	author = {{de Kleuver, Joost} and {Bronzwaer, Thomas} and {Falcke, Heino} and {Narayan, Ramesh} and {Mizuno, Yosuke} and {Porth, Oliver} and {Olivares, Hector}},
	title = {Testing the existence of event horizons against rotating reflecting surfaces},
	DOI= "10.1051/0004-6361/202348489",
	url= "https://doi.org/10.1051/0004-6361/202348489",
	journal = {Astronomy and Astrophysics},
	year = 2024,
	volume = 689,
	pages = "A197",
}

@article{Olivares2020,
    author = {Olivares, Hector and Younsi, Ziri and Fromm, Christian M and De Laurentis, Mariafelicia and Porth, Oliver and Mizuno, Yosuke and Falcke, Heino and Kramer, Michael and Rezzolla, Luciano},
    title = {How to tell an accreting boson star from a black hole},
    journal = {Monthly Notices of the Royal Astronomical Society},
    volume = {497},
    number = {1},
    pages = {521-535},
    year = {2020},
    month = {09},
    issn = {0035-8711},
    doi = {10.1093/mnras/staa1878},
    url = {https://doi.org/10.1093/mnras/staa1878},
}

@article{Gimeno_2021,
  title = {Magnetized accretion disks around Kerr black holes with scalar hair: Nonconstant angular momentum disks},
  author = {Gimeno-Soler, Sergio and Font, Jos{\'e} A. and Herdeiro, Carlos and Radu, Eugen},
  journal = {Phys. Rev. D},
  volume = {104},
  issue = {10},
  pages = {103008},
  numpages = {21},
  year = {2021},
  month = {Nov},
  publisher = {American Physical Society},
  doi = {10.1103/PhysRevD.104.103008},
  url = {https://link.aps.org/doi/10.1103/PhysRevD.104.103008}
}

@article{Gimeno_2019,
  title = {Magnetized accretion disks around Kerr black holes with scalar hair: Constant angular momentum disks},
  author = {Gimeno-Soler, Sergio and Font, Jos{\'e} A. and Herdeiro, Carlos and Radu, Eugen},
  journal = {Phys. Rev. D},
  volume = {99},
  issue = {4},
  pages = {043002},
  numpages = {19},
  year = {2019},
  month = {Feb},
  publisher = {American Physical Society},
  doi = {10.1103/PhysRevD.99.043002},
  url = {https://link.aps.org/doi/10.1103/PhysRevD.99.043002}
}

@article{Universal,
  title = {Universal polarimetric signatures of the black hole photon ring},
  author = {Himwich, Elizabeth and Johnson, Michael D. and Lupsasca, Alexandru and Strominger, Andrew},
  journal = {Phys. Rev. D},
  volume = {101},
  issue = {8},
  pages = {084020},
  numpages = {13},
  year = {2020},
  month = {Apr},
  publisher = {American Physical Society},
  doi = {10.1103/PhysRevD.101.084020},
  url = {https://link.aps.org/doi/10.1103/PhysRevD.101.084020}
}

@article{Deliyski2026,
  title = {{P}olarized equatorial emission around {K}err black holes with synchronized scalar hair. {I}. {D}irect images},
  volume = {86},
  ISSN = {1434-6052},
  url = {http://dx.doi.org/10.1140/epjc/s10052-026-16103-2},
  DOI = {10.1140/epjc/s10052-026-16103-2},
  number = {7},
  journal = {The European Physical Journal C},
  publisher = {Springer Science and Business Media LLC},
  author = {Deliyski,  Valentin O. and Gyulchev,  Galin N. and Doneva,  Daniela D. and Nedkova,  Petya G. and Yazadjiev,  Stoytcho S.},
  year = {2026},
  month = July 
}

@article{hairysol1,
  title={Stationary scalar clouds around rotating black holes},
  author={Hod, Shahar},
  journal={Physical Review D},
  volume={86},
  number={10},
  pages={104026},
  year={2012},
  publisher={APS},
  DOI={https://doi.org/10.1103/PhysRevD.86.104026}
}

@article{herdeiro2015,
  title={Construction and physical properties of Kerr black holes with scalar hair},
  author={Herdeiro, Carlos and Radu, Eugen},
  journal={Classical and Quantum Gravity},
  volume={32},
  number={14},
  pages={144001},
  year={2015},
  doi = {10.1088/0264-9381/32/14/144001},
  publisher={IOP Publishing}
}

@article{herdeiro2014,
  title={Kerr black holes with scalar hair},
  author={Herdeiro, Carlos AR and Radu, Eugen},
  journal={Physical review letters},
  volume={112},
  number={22},
  pages={221101},
  year={2014},
  doi = {https://doi.org/10.1103/PhysRevLett.112.221101},
  publisher={APS}
}

@article{collodel2020rotating,
  title={Rotating tensor-multiscalar black holes with two scalars},
  author={Collodel, Lucas G and Doneva, Daniela D and Yazadjiev, Stoytcho S},
  journal={Physical Review D},
  volume={102},
  number={8},
  pages={084032},
  year={2020},
  publisher={APS},
  doi = {https://doi.org/10.1103/PhysRevD.102.084032},
}

@article{Broderick_2023,
doi = {10.3847/1538-4357/acf9f6},
url = {https://doi.org/10.3847/1538-4357/acf9f6},
year = {2023},
month = {nov},
publisher = {The American Astronomical Society},
volume = {958},
number = {2},
pages = {114},
author = {Broderick, Avery E. and Salehi, Kiana and Georgiev, Boris},
title = {Shadow Implications: What Does Measuring the Photon Ring Imply for Gravity?},
journal = {The Astrophysical Journal}
}

@article{Staelens_2023,
  title = {Black hole photon rings beyond general relativity},
  author = {Staelens, Seppe and Mayerson, Daniel R. and Bacchini, Fabio and Ripperda, Bart and K{\"u}chler, Lorenzo},
  journal = {Phys. Rev. D},
  volume = {107},
  issue = {12},
  pages = {124026},
  numpages = {30},
  year = {2023},
  month = {Jun},
  publisher = {American Physical Society},
  doi = {10.1103/PhysRevD.107.124026},
  url = {https://link.aps.org/doi/10.1103/PhysRevD.107.124026}
}

@article{Paugnat_2022,
	author = {{Paugnat, H.} and {Lupsasca, A.} and {Vincent, F. H.} and {Wielgus, M.}},
	title = {Photon ring test of the Kerr hypothesis: Variation in the ring shape},
	DOI= "10.1051/0004-6361/202244216",
	url= "https://doi.org/10.1051/0004-6361/202244216",
	journal = {Astronomy and Astrophysics},
	year = 2022,
	volume = 668,
	pages = "A11",
}

@Article{Tiede_2022,
AUTHOR = {Tiede, Paul and Johnson, Michael D. and Pesce, Dominic W. and Palumbo, Daniel C. M. and Chang, Dominic O. and Galison, Peter},
TITLE = {Measuring Photon Rings with the ngEHT},
JOURNAL = {Galaxies},
VOLUME = {10},
YEAR = {2022},
NUMBER = {6},
ARTICLE-NUMBER = {111},
URL = {https://www.mdpi.com/2075-4434/10/6/111},
ISSN = {2075-4434},
DOI = {10.3390/galaxies10060111}
}

@article{Shavelle_2024,
doi = {10.3847/2041-8213/ad6000},
url = {https://doi.org/10.3847/2041-8213/ad6000},
year = {2024},
month = {jul},
publisher = {The American Astronomical Society},
volume = {970},
number = {1},
pages = {L24},
author = {Shavelle, Kaitlyn M. and Palumbo, Daniel C. M.},
title = {Prospects for the {D}etection of the {S}gr {A}* {P}hoton {R}ing with {N}ext-generation {E}vent {H}orizon {T}elescope {P}olarimetry},
journal = {The Astrophysical Journal Letters}
}

@article{Moscibrodzka_2021,
    author = {Mo{\'s}cibrodzka, M and Janiuk, A and De Laurentis, M},
    title = {Unraveling circular polarimetric images of magnetically arrested accretion flows near event horizon of a black hole},
    journal = {Monthly Notices of the Royal Astronomical Society},
    volume = {508},
    number = {3},
    pages = {4282-4296},
    year = {2021},
    month = {12},
    issn = {0035-8711},
    doi = {10.1093/mnras/stab2790},
    url = {https://doi.org/10.1093/mnras/stab2790},
    eprint = {https://academic.oup.com/mnras/article-pdf/508/3/4282/40825131/stab2790.pdf},
}

@article{Tamar_2024,
doi = {10.3847/1538-4357/ad8dd5},
url = {https://doi.org/10.3847/1538-4357/ad8dd5},
year = {2024},
month = {dec},
publisher = {The American Astronomical Society},
volume = {977},
number = {2},
pages = {147},
author = {Tamar, Aditya and Palumbo, Daniel C. M.},
title = {{P}hoton {R}ing {P}olarimetry with {N}ext-generation {B}lack {H}ole {I}maging. {I}. {M}87*},
journal = {The Astrophysical Journal}
}

@article{PhysRevD.107.043030,
  title = {Adaptive analytical ray tracing of black hole photon rings},
  author = {C{\'a}rdenas-Avenda{\~n}o, Alejandro and Lupsasca, Alexandru and Zhu, Hengrui},
  journal = {Phys. Rev. D},
  volume = {107},
  issue = {4},
  pages = {043030},
  numpages = {44},
  year = {2023},
  month = {Feb},
  publisher = {American Physical Society},
  doi = {10.1103/PhysRevD.107.043030},
  url = {https://link.aps.org/doi/10.1103/PhysRevD.107.043030}
}

@Article{Johnson_2023,
AUTHOR = {Johnson, Michael D. and Akiyama, Kazunori and Blackburn, Lindy and Bouman, Katherine L. and Broderick, Avery E. and Cardoso, Vitor and Fender, Rob P. and Fromm, Christian M. and Galison, Peter and G{\'o}mez, José L. and Haggard, Daryl and Lister, Matthew L. and Lobanov, Andrei P. and Markoff, Sera and Narayan, Ramesh and Natarajan, Priyamvada and Nichols, Tiffany and Pesce, Dominic W. and Younsi, Ziri and Chael, Andrew and Chatterjee, Koushik and Chaves, Ryan and Doboszewski, Juliusz and Dodson, Richard and Doeleman, Sheperd S. and Elder, Jamee and Fitzpatrick, Garret and Haworth, Kari and Houston, Janice and Issaoun, Sara and Kovalev, Yuri Y. and Levis, Aviad and Lico, Rocco and Marcoci, Alexandru and Martens, Niels C. M. and Nagar, Neil M. and Oppenheimer, Aaron and Palumbo, Daniel C. M. and Ricarte, Angelo and Rioja, María J. and Roelofs, Freek and Thresher, Ann C. and Tiede, Paul and Weintroub, Jonathan and Wielgus, Maciek},
TITLE = {Key Science Goals for the Next-Generation Event Horizon Telescope},
JOURNAL = {Galaxies},
VOLUME = {11},
YEAR = {2023},
NUMBER = {3},
ARTICLE-NUMBER = {61},
URL = {https://www.mdpi.com/2075-4434/11/3/61},
ISSN = {2075-4434},
DOI = {10.3390/galaxies11030061}
}

@article{Vincent_2016,
  title = {Astrophysical imaging of Kerr black holes with scalar hair},
  author = {Vincent, F. H. and Gourgoulhon, E. and Herdeiro, C. and Radu, E.},
  journal = {Phys. Rev. D},
  volume = {94},
  issue = {8},
  pages = {084045},
  numpages = {12},
  year = {2016},
  month = {Oct},
  publisher = {American Physical Society},
  doi = {10.1103/PhysRevD.94.084045},
  url = {https://link.aps.org/doi/10.1103/PhysRevD.94.084045}
}

@article{Galin_2026,
  title = {Images of the thin accretion disk around Kerr black holes coupled to time periodic scalar fields},
  author = {Gyulchev, Galin N. and Doneva, Daniela D. and Deliyski, Valentin O. and Nedkova, Petya G. and Yazadjiev, Stoytcho S.},
  journal = {Phys. Rev. D},
  volume = {114},
  issue = {2},
  pages = {024074},
  numpages = {31},
  year = {2026},
  month = {Jul},
  publisher = {American Physical Society},
  doi = {10.1103/3xwd-4p4w},
  url = {https://link.aps.org/doi/10.1103/3xwd-4p4w}
}

@article{Narayan_2021,
doi = {10.3847/1538-4357/abf117},
url = {https://doi.org/10.3847/1538-4357/abf117},
year = {2021},
month = {may},
publisher = {The American Astronomical Society},
volume = {912},
number = {1},
pages = {35},
author = {Narayan, Ramesh and Palumbo, Daniel C. M. and Johnson, Michael D. and Gelles, Zachary and Himwich, Elizabeth and Chang, Dominic O. and Ricarte, Angelo and Dexter, Jason and Gammie, Charles F. and Chael, Andrew A. and {The Event Horizon Telescope Collaboration}},
title = {The Polarized Image of a Synchrotron-emitting Ring of Gas Orbiting a Black Hole},
journal = {The Astrophysical Journal}
}

@article{Gauss_Bonnet_pol,
	author = {Qin, Xin and Chen, Songbai and Jing, Jiliang},
	title = {{Polarized image of an equatorial emitting ring around a 4D Gauss-Bonnet black hole}},
	doi = {10.1140/epjc/s10052-022-10753-8},
	url= {https://doi.org/10.1140/epjc/s10052-022-10753-8},
	journal = {Eur. Phys. J. C},
	year = 2022,
	volume = 82,
	number = 9,
	pages = "784",
}

@article{Marszewski_2021,
doi = {10.3847/1538-4357/ac1b28},
url = {https://doi.org/10.3847/1538-4357/ac1b28},
year = {2021},
month = {oct},
publisher = {The American Astronomical Society},
volume = {921},
number = {1},
pages = {17},
author = {Marszewski, Andrew and Prather, Ben S. and Joshi, Abhishek V. and Pandya, Alex and Gammie, Charles F.},
title = {Updated Transfer Coefficients for Magnetized Plasmas},
journal = {The Astrophysical Journal}
}

@article{Wormhole_paper,
  title = {Polarized image of equatorial emission in horizonless spacetimes: Traversable wormholes},
  author = {Deliyski, Valentin and Gyulchev, Galin and Nedkova, Petya and Yazadjiev, Stoytcho},
  journal = {Phys. Rev. D},
  volume = {106},
  issue = {10},
  pages = {104024},
  numpages = {24},
  year = {2022},
  month = {Nov},
  publisher = {American Physical Society},
  doi = {10.1103/PhysRevD.106.104024},
  url = {https://link.aps.org/doi/10.1103/PhysRevD.106.104024}
}

@ARTICLE{Mahadevan_1996,
       author = {{Mahadevan}, Rohan and {Narayan}, Ramesh and {Yi}, Insu},
        title = "{Harmony in Electrons: Cyclotron and Synchrotron Emission by Thermal Electrons in a Magnetic Field}",
      journal = {\apj},
         year = 1996,
        month = jul,
       volume = {465},
        pages = {327},
          doi = {10.1086/177422},
 primaryClass = {astro-ph},
       adsurl = {https://ui.adsabs.harvard.edu/abs/1996ApJ...465..327M}
}

@article{Naked_singularity_pol,
  title = {Polarized image of equatorial emission in horizonless spacetimes: Naked singularities},
  author = {Deliyski, Valentin and Gyulchev, Galin and Nedkova, Petya and Yazadjiev, Stoytcho},
  journal = {Phys. Rev. D},
  volume = {108},
  issue = {10},
  pages = {104049},
  numpages = {20},
  year = {2023},
  month = {Nov},
  publisher = {American Physical Society},
  doi = {10.1103/PhysRevD.108.104049},
  url = {https://link.aps.org/doi/10.1103/PhysRevD.108.104049}
}

@ARTICLE{Tsanimir_2025,
       author = {{Angelov}, Tsanimir and {Bekir}, Rasim and {Gyulchev}, Galin and {Nedkova}, Petya and {Yazadjiev}, Stoytcho},
        title = "{Polarized equatorial emission and hot spots around black holes with a dark matter halo}",
      journal = {European Physical Journal C},
         year = 2025,
        month = sep,
       volume = {85},
       number = {9},
          eid = {1075},
        pages = {1075},
          doi = {10.1140/epjc/s10052-025-14537-8},
 primaryClass = {gr-qc},
       adsurl = {https://ui.adsabs.harvard.edu/abs/2025EPJC...85.1075A}
}

@article{Gyulchev2021,
  title = {{Image of the thin accretion disk around compact objects in the Einstein-Gauss-Bonnet gravity}},
  volume = {81},
  ISSN = {1434-6052},
  url = {http://dx.doi.org/10.1140/epjc/s10052-021-09624-5},
  DOI = {10.1140/epjc/s10052-021-09624-5},
  number = {10},
  journal = {The European Physical Journal C},
  publisher = {Springer Science and Business Media LLC},
  author = {Gyulchev,  Galin and Nedkova,  Petya and Vetsov,  Tsvetan and Yazadjiev,  Stoytcho},
  year = {2021},
  month = Oct 
}

@article{Gyulchev2019,
  title = {{Image of the Janis-Newman-Winicour naked singularity with a thin accretion disk}},
  author = {Gyulchev, Galin and Nedkova, Petya and Vetsov, Tsvetan and Yazadjiev, Stoytcho},
  journal = {Phys. Rev. D},
  volume = {100},
  issue = {2},
  pages = {024055},
  numpages = {13},
  year = {2019},
  month = {Jul},
  publisher = {American Physical Society},
  doi = {10.1103/PhysRevD.100.024055},
  url = {https://link.aps.org/doi/10.1103/PhysRevD.100.024055}
}

@article{Gyulchev2020,
  title = {Observational signatures of strongly naked singularities: image of the thin accretion disk},
  volume = {80},
  ISSN = {1434-6052},
  url = {http://dx.doi.org/10.1140/epjc/s10052-020-08575-7},
  DOI = {10.1140/epjc/s10052-020-08575-7},
  number = {11},
  journal = {The European Physical Journal C},
  publisher = {Springer Science and Business Media LLC},
  author = {Gyulchev,  Galin and Kunz,  Jutta and Nedkova,  Petya and Vetsov,  Tsvetan and Yazadjiev,  Stoytcho},
  year = {2020},
  month = Nov 
}

@article{Paul_2020,
doi = {10.1088/1475-7516/2020/03/055},
url = {https://doi.org/10.1088/1475-7516/2020/03/055},
year = {2020},
month = {mar},
publisher = {},
volume = {2020},
number = {03},
pages = {055},
author = {Paul, Suvankar and Shaikh, Rajibul and Banerjee, Pritam and Sarkar, Tapobrata},
title = {Observational signatures of wormholes with thin accretion disks},
journal = {Journal of Cosmology and Astroparticle Physics}
}

@article{Rahaman2021,
title = {Thin accretion disks around traversable wormholes},
journal = {Nuclear Physics B},
volume = {972},
pages = {115548},
year = {2021},
issn = {0550-3213},
doi = {https://doi.org/10.1016/j.nuclphysb.2021.115548},
url = {https://www.sciencedirect.com/science/article/pii/S0550321321002455},
author = {Farook Rahaman and Tuhina Manna and Rajibul Shaikh and Somi Aktar and Monimala Mondal and Bidisha Samanta}
}

@article{Wan2026,
  title = {Distorting Kerr images with parity-odd scalar hair},
  author = {Wan, Qian and Hou, Yehui and Huang, Yang and Li, Peng-Cheng and Guo, Minyong and Chen, Bin},
  journal = {Phys. Rev. D},
  volume = {114},
  issue = {8},
  pages = {084012},
  numpages = {12},
  year = {2026},
  month = {Oct},
  publisher = {American Physical Society},
  doi = {10.1103/f3wb-bv6m},
  url = {https://link.aps.org/doi/10.1103/f3wb-bv6m}
}

@article{Cunha2015,
  title = {Shadows of Kerr Black Holes with Scalar Hair},
  author = {Cunha, Pedro V. P. and Herdeiro, Carlos A. R. and Radu, Eugen and R{\'u}narsson, Helgi F.},
  journal = {Phys. Rev. Lett.},
  volume = {115},
  issue = {21},
  pages = {211102},
  numpages = {5},
  year = {2015},
  month = {Nov},
  publisher = {American Physical Society},
  doi = {10.1103/PhysRevLett.115.211102},
  url = {https://link.aps.org/doi/10.1103/PhysRevLett.115.211102}
}

@article{Gyulchev2024,
  title = {Shadows of rotating hairy Kerr black holes coupled to time periodic scalar fields with a nonflat target space},
  author = {Gyulchev, Galin N. and Roy, Ayush and Collodel, Lucas G. and Nedkova, Petya G. and Yazadjiev, Stoytcho S. and Doneva, Daniela D.},
  journal = {Phys. Rev. D},
  volume = {109},
  issue = {10},
  pages = {104051},
  numpages = {20},
  year = {2024},
  month = {May},
  publisher = {American Physical Society},
  doi = {10.1103/PhysRevD.109.104051},
  url = {https://link.aps.org/doi/10.1103/PhysRevD.109.104051}
}

@article{M_Johnson_2021,
  title = {Polarized image of equatorial emission in the Kerr geometry},
  author = {Gelles, Zachary and Himwich, Elizabeth and Johnson, Michael D. and Palumbo, Daniel C. M.},
  journal = {Phys. Rev. D},
  volume = {104},
  issue = {4},
  pages = {044060},
  numpages = {25},
  year = {2021},
  month = {Aug},
  publisher = {American Physical Society},
  doi = {10.1103/PhysRevD.104.044060},
  url = {https://link.aps.org/doi/10.1103/PhysRevD.104.044060}
}

@article{EHT_M87_VII,
doi = {10.3847/2041-8213/abe4de},
url = {https://doi.org/10.3847/2041-8213/abe4de},
year = {2021},
month = {mar},
publisher = {The American Astronomical Society},
volume = {910},
number = {1},
pages = {L13},
author = {{The Event Horizon Telescope Collaboration}},
title = {First {M}87 {E}vent {H}orizon {T}elescope {R}esults. {VIII}. {M}agnetic {F}ield {S}tructure near {T}he {E}vent {H}orizon},
journal = {The Astrophysical Journal Letters}
}

@article{EHT_sgrA_VIII,
doi = {10.3847/2041-8213/ad2df1},
url = {https://doi.org/10.3847/2041-8213/ad2df1},
year = {2024},
month = {mar},
publisher = {The American Astronomical Society},
volume = {964},
number = {2},
pages = {L26},
journal = {The Astrophysical Journal Letters},
title = {{F}irst {S}agittarius {A}* {E}vent {H}orizon {T}elescope {R}esults. {VIII}. {P}hysical {I}nterpretation of the {P}olarized {R}ing},
author = {{The Event Horizon Telescope Collaboration}}
}

@article{EHT_SgrA_I,
doi = {10.3847/2041-8213/ac6674},
url = {https://doi.org/10.3847/2041-8213/ac6674},
year = {2022},
month = {may},
publisher = {The American Astronomical Society},
volume = {930},
number = {2},
pages = {L12},
author = {{Event Horizon Telescope Collaboration}},
title = {{F}irst {S}agittarius {A}* {E}vent {H}orizon {T}elescope {R}esults. {I}. {T}he {S}hadow of the {S}upermassive {B}lack {H}ole in the {C}enter of the {M}ilky {W}ay},
journal = {The Astrophysical Journal Letters}
}

@article{EHT_M87_I,
doi = {10.3847/2041-8213/ab0ec7},
url = {https://doi.org/10.3847/2041-8213/ab0ec7},
year = {2019},
month = {apr},
publisher = {The American Astronomical Society},
volume = {875},
number = {1},
pages = {L1},
author = {{The Event Horizon Telescope Collaboration}},
title = {{F}irst {M}87 {E}vent {H}orizon {T}elescope {R}esults. {I}. {T}he {S}hadow of the {S}upermassive {B}lack {H}ole},
journal = {The Astrophysical Journal Letters}
}

@article{Akiyama_2019,
doi = {10.3847/2041-8213/ab1141},
url = {https://doi.org/10.3847/2041-8213/ab1141},
year = {2019},
month = {apr},
publisher = {The American Astronomical Society},
volume = {875},
number = {1},
pages = {L6},
author = {{The Event Horizon Telescope Collaboration}},
title = {{First M87 Event Horizon Telescope Results. VI. The Shadow and Mass of the Central Black Hole}},
journal = {The Astrophysical Journal Letters}
}

\end{document}